# Conservative Rupture Force Relating Co-seismic Rupture and Afterslip Process

**Yuxuan Hou[1], Zeyan Zhao[2], Han Yue*[1]**

1. State Key Laboratory of Earthquake Dynamics and Forecasting, SESS, Peking University, Beijing 100871, China

2. State Key Laboratory of Earthquake Dynamics and Forecasting, Institute of Geology, China Earthquake Administration, Beijing 100029, China

## Key Points

- We derive an equilibrium slip distribution (equilibrium state), which is maintained throughout the earthquake cycle.
- We propose and validate a "conservative rupture force" (CRF) that links co-seismic rupture to afterslip.
- We apply the CRF calculation to the 2015 Gorkha earthquake and evaluate the afterslip release process.

## Abstract

The post-seismic phase plays a crucial role in the earthquake cycle by releasing co-seismic stress and reshaping the fault-zone stress field. Traditionally, the relationship between afterslip and co-seismic slip is made by the ratio of their respective moment ($R_M$). However, in this study, we perform numerical and theoretical analyses to reveal that $R_M$ is inherently unstable. To quantify the afterslip, we first introduce an inter-seismic "equilibrium state" as a reference from which the pure afterslip is separated, while the afterslip releases the co-seismic stress loading over the velocity-strengthening region after correcting for this equilibrium state. To define the relationship between co- and afterslips, we propose a novel physical quantity termed the "conservative rupture force" (CRF), which is defined as the area integration of co-seismic stress drop. The CRF is equal to the area integration of stress loading over an infinite fault plane as well as the area integration of afterslip stress release over an infinite time and fault plane. This relationship holds for both full-space and half-space faults. We derive a dimensionless parameter as the ratio between afterslip and co-seismic CRFs as $R_F$. We also propose a strategy to derive such a factor with given co-seismic and afterslip models. We demonstrate that after correcting for the spatial partitioning of stress release, $R_F$ converges to one in an infinite time. This strategy is applied to the 2015 Mw 7.8 Gorkha earthquake, and we find that afterslips released 30% of co-seismic stress loading over 5.5 years after the mainshock, while a large portion of the fault plane remained locked. The introduction of the equilibrium state and CRF offers a fresh quantitative view to understand relationships between inter-, co- and post-seismic fault slips.

## Plain Language Summary

When an earthquake occurs, the sudden slip on a fault (the mainshock) does not end the story. In months to years following that, the fault continues to slip slowly and quietly—a process called afterslip. Afterslip can release stress, trigger aftershocks, and even influence future earthquakes. But how large is afterslip compared to the mainshock? Traditionally, scientists compare their seismic moments (a measure of released energy) using a ratio called $\mathrm{R_M}$. However, this study shows that $R_M$ is unreliable: it depends strongly on how large we assume the fault to be, rather than reflecting a true physical relationship. To overcome this problem, we introduce a new physical quantity called the "conservative rupture force" (CRF). Instead of looking at energy (moment), CRF looks at the total force exerted by stress changes over the fault area. Because the fault also slides slowly as part of normal plate motion, and this background sliding overlaps with the afterslip, we first separate the afterslip from this background before measuring it. Applying our method to the 2015 Mw 7.8 Gorkha earthquake, we find that afterslip released about 30% of the co-seismic CRF over 5.5 years. This new framework provides a more stable and meaningful way to compare afterslip across different earthquakes, helping us to better understand how faults behave between major ruptures.

## Introduction

An earthquake cycle is commonly described as three consecutive stages: long-term inter-seismic loading, co-seismic rupture and post-seismic adjustment (Wang et al., 2012; Avouac, 2015; Govers et al., 2018). The post-seismic phase plays a crucial role, as it releases co-seismic stress, reshapes the fault-zone stress field, links successive earthquakes and influences future rupture patterns. Post-seismic deformation is mainly driven by three mechanisms: poroelastic rebound, viscoelastic relaxation, and afterslip. Poroelastic rebound resulting from fluid migration driven by co-seismic stress loading typically dominates at the fault tip and over a short time scale (Peltzer et al., 1998; Jónsson et al., 2003). In contrast, both afterslip and viscoelastic relaxation result in deformation that covers a larger spatial area and lasts longer (Deng et al., 1998; Bürgmann & Dresen, 2008; Avouac, 2015). Traditionally, afterslip is regarded as the early, relatively fast response, whereas viscoelastic relaxation is viewed as the longer-term, slower response of the ductile lower crust and upper mantle (Diao et al., 2014; Reilinger et al., 2000). However, this conventional "early afterslip, later viscoelasticity" sequence is not universal: in several subduction and continental-thrust environments, viscoelastic relaxation includes a strong transient component that emerges within months of the mainshock, so that the relative timing and amplitude of afterslip and viscoelastic relaxation vary considerably among megathrusts (e.g., Sun & Wang, 2015; Periollat et al., 2022). Afterslip-related deformation of strong earthquakes (Mw>6) is commonly detectable using modern geodetic techniques (e.g., GNSS, InSAR), offering critical insights into fault frictional properties and the rheological behaviors of the seismogenic zone (e.g., Hsu et al., 2006; Zhao & Yue, 2023).

The afterslip process can be described by the steady-state rate-and-state friction mechanism (ss-RSF) (Marone et al., 1991; Scholz, 1998):

$$\tau = \sigma(f_0 + (a - b) ln \frac{v}{v_0}) \quad (1)$$

where $\tau$ represents the shear stress acting on the fault, $\sigma$ is the effective normal stress, $f_0$ is the friction coefficient when slip velocity ($v$) equals the reference velocity ($v_0$), $a$ and $b$ are non-dimensional frictional parameters. The sign of the parameter $(a - b)$ determines the velocity dependence of friction: negative $(a - b)$ corresponds to velocity weakening (VW) behavior, which promotes unstable slip (earthquake rupture), whereas positive $(a - b)$ corresponds to velocity strengthening (VS) behavior, which favors stable fault slip (Gu et al., 1984; Ruina, 1983). A widely adopted fault-zone model posits that VW and VS zones alternate in distribution with depth, as illustrated in Figure 1a. Typically, seismogenic fault segments exhibit negative $(a - b)$ values, while creeping sections show positive $(a - b)$.

Based on this picture, a simplified earthquake cycle can be described as follows: during the inter-seismic period, the locked VW zone is loaded by the creeping of the VS zone. Once this loading reaches a certain threshold, earthquake rupture (failure) occurs in the VW zone (near the VW-VS transition), which imposes co-seismic stress loading on the surrounding VS zone; the stress loading of co-seismic rupture thereby drives afterslip (Perfettini et al., 2018). In the following post-seismic period, afterslip releases the stress loading and converges to an inter-seismic creeping state (Figure 1b), initiating a new earthquake cycle (Scholz, 1998; Avouac, 2015). During this process, afterslip significantly alters the post-seismic stress distribution in the crust and may drive the occurrence of aftershock sequences (Bürgmann et al., 2002; Perfettini & Avouac, 2004).

Despite its importance, the first-order characteristic (i.e., magnitude) of the afterslip phenomenon remains poorly understood. Intuitively, the magnitude of afterslip should scale with the mainshock magnitude, since a larger mainshock commonly loads larger source areas. This suggests quantifying both processes by their seismic moments, which yields the dimensionless ratio $R_M$ defined as:

$$R_M = \frac{M_{aft}}{M_{co}} \quad (2)$$

where $M_{aft}$ is the afterslip moment and $M_{co}$ is the co-seismic moment. However, global and local estimations of $R_M$ values failed to establish a convincing relationship over different earthquakes (Wimpenny et al., 2017; Alwahedi & Hawthorne, 2019), and the statistical relationships between $R_M$ and mainshock rupture characteristics (e.g., rake, depth, and rupture aspect ratio) are not significant (Churchill et al., 2022). Churchill et al. (2022) show that the moments of afterslip models could be as small as 0.4% or as large as 4 times the co-seismic moment, thus introducing about three orders of magnitude variation. Even for the same mainshock, the estimations of $R_M$ can vary by more than one order of magnitude across different studies (Churchill et al., 2022). For instance, studies of the 2004 Mw 9.1 Sumatra earthquake report an $R_M$ of 3.4 (Panet et al., 2010) and 55.4 (Hashimoto et al., 2006), differing by a factor of approximately 16. Similarly, for the 2008 Mw 8.0 Sichuan earthquake, $R_M$ estimates range from 0.4 (Shao et al., 2011) to 13.4 (Diao et al., 2018), a difference of about a factor of 34.

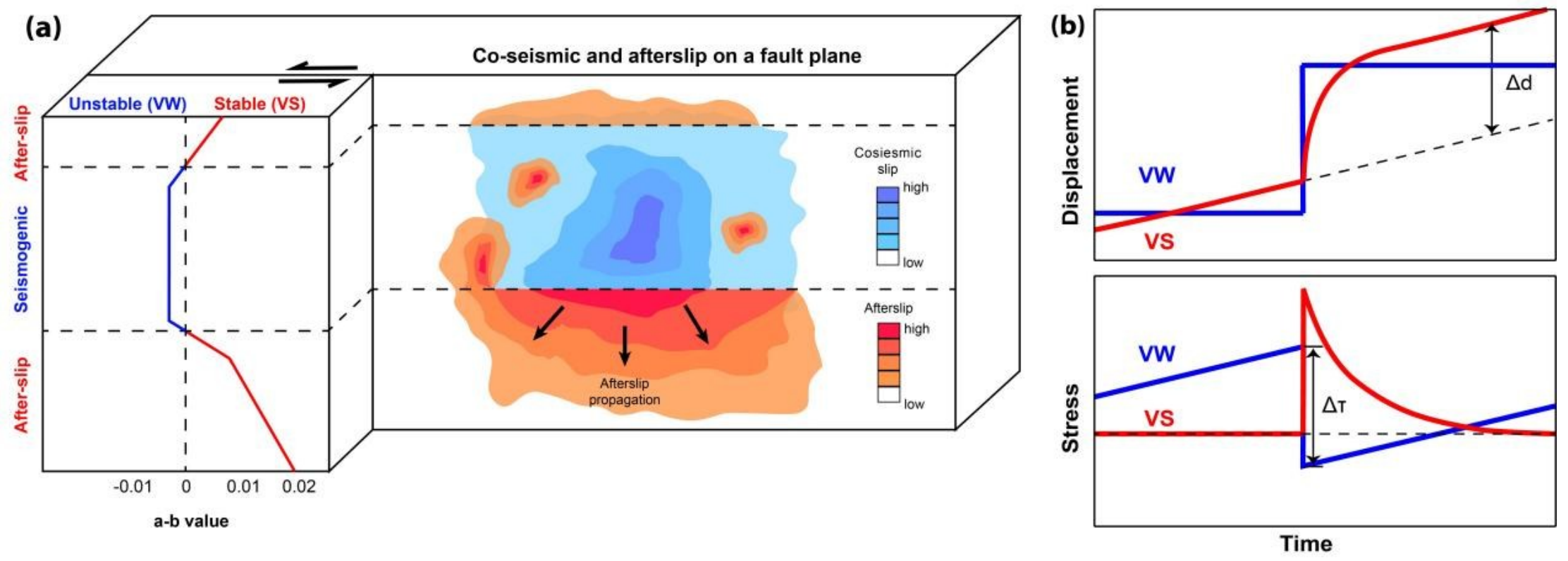

**Figure 1.** An illustrative model for co- and afterslips. (a) A representative depth profile for the $a-b$ parameter (Blanpied et al., 1991) is plotted on the left. VS and VW zones are shown in red and blue, respectively. Co-seismic slip and afterslip distribution on an idealized strike-slip fault is plotted on the right (modified from Churchill et al. [2022]). (b) Theoretical evolution of displacement (top) and stress (bottom) in the VS (red) and VW (blue) zones. $\Delta d$ denotes the pure afterslip when background creep is removed, and $\Delta\tau$ denotes the co-seismic stress drop.

The significant variation of $R_M$ indicates the possible impacts from data processing and inversion strategy or that from inappropriate selection of physical parameters. For an individual event, the co-seismic slips could be constrained by both global seismic networks as well as regional geodetic observations (e.g. GPS and InSAR), which provide strong constraints to the earthquake moment. However, afterslip models can only be constrained by geodetic data, introducing stronger variations from data coverage and time window selection. Thus the reported one-order-of-magnitude variation of $R_M$ may originate from afterslip models. However, the uncertainty of slip models (10 times) may not explain the reported $R_M$ variation (1000 times) over global events. This discrepancy emphasizes the importance of physical parameter selection: whether $R_M$ is an appropriate measurement for the relative afterslip magnitude.

In this study, we demonstrate the instability of $R_M$ and seek to establish a robust alternative parameter. Firstly, we demonstrate that afterslip tends to release the co-seismic stress loading in the VS area, thus maintaining a force balance (area integration of stress loading) instead of a moment balance. In the simulation, an equilibrium state is introduced to make the necessary correction for inter-seismic loading. Secondly, we demonstrate that $R_M$ is an unstable metric by numerical and theoretical analysis. Subsequently, we introduce a new term "conservative rupture force" (CRF), which is a conservative quantity to link co- and afterslips. A ratio between CRFs released by afterslip and co-seismic slips ($R_F$) is thus proposed, which theoretically converges to 1 in an ideal environment. Finally, we apply this quantity to the 2015 Mw 7.8 Gorkha, Nepal earthquake, to demonstrate how to calculate and correct for $R_F$ in real co-seismic and afterslip models.

## 2 Methods, Derivations & Synthetic Tests

### 2.1 Numerical Model

In the inversion results of real afterslips, the slip models are subject to various uncertainties from observation and modeling errors, fault parameterization and data coverage, making it difficult to estimate the physical mechanism causing $R_M$ variations. In this study, we rely on numerical simulation, which is performed with known fault geometry and physical processes, as well as precisely measured co- and afterslips. The numerical approach removes the observational and inversion uncertainties and provides clean cases to evaluate afterslip behaviors under ideal conditions. Following this logic, we perform quasi-static numerical simulations of a single planar fault based on the boundary integral method (BIM) with ss-RSF (Zhao & Yue, 2023). The fault plane is discretized into $N$ square subfaults of 1 km × 1 km (Table 1); the total fault-plane dimension is given separately for each test (Sections 2.3 and 2.4), while the elastic medium is a homogeneous, isotropic half-space, and the shear-stress kernels $\mathbf{G}_{ij}$ (the shear stress change on subfault $i$ caused by unit slip on subfault $j$) are computed from the analytical solution for a rectangular dislocation (Okada, 1992). For a planar fault in a homogeneous medium, it is a reasonable assumption that the normal stress changes on subfaults are negligible in comparison with the shear stress changes. The governing equations are:

$$\Delta\tau_i(t) = \sigma_i(a-b)_i ln\frac{v_i(t+\Delta t)}{v_i(t)} \tag{3}$$

$$\Delta\tau_i(t) = \sum_{j=1}^{N}\mathbf{G}_{ij}\left[v_j(t) - v_{pl}\right]\Delta t \tag{4}$$

where $i, j$ index the subfaults ($i, j = 1, \ldots, N$), $\Delta t$ is the time step, and $v_{pl}$ is the background loading rate. For simplicity, we assume that the normal stress $\sigma$ and frictional parameters $a$ and $b$ are uniform across the fault plane. At $t = 0$ the fault is at the inter-seismic equilibrium state ($v_i(0) = v_i^{equ}$; Section 2.2), onto which the instantaneous co-seismic stress change $\Delta\tau_i(0)$ is superimposed; the initial slip velocities are obtained from Equation 3. Tectonic loading at rate $v_{pl}$ continues throughout the afterslip evolution. All parameters are listed in Table 1. Since in the ss-RSF formulation the effective normal stress $\sigma$ and the frictional parameters $a$ and $b$ never enter the evolution equations separately, but only through the combined product $\sigma(a-b)$, this single quantity therefore controls the evolution and is the value listed in Table 1.

Because we focus on the slow post-seismic evolution, inertia and radiation damping are neglected, and the co-seismic phase is not simulated dynamically: the co-seismic slip (and hence the co-seismic stress change) is prescribed analytically from a prescribed slip/stress-drop pattern. The VW (co-seismic) patch is held locked during the post-seismic phase, and only the surrounding VS region evolves under ss-RSF. We use the steady-state form of RSF (Equation 1) rather than the full rate-and-state law, because we only care about the long term afterslip evolution, especially its final state in infinite time. Since the concerned time scale ( at least years) is much longer than the state-relaxation time $D_c/v$, the frictional response is quasi-static (Marone et al., 1991); this makes the stress-velocity relation single-valued (Equation 1), which is also the premise of the equilibrium state derivation (Section 2.2). A limitation is that transient strengthening/weakening during rapid velocity changes is not captured, which is acceptable for the slow afterslip considered here. In our simulation, we aim to obtain the relationship of co- and afterslips in the simplest setting, which is described by afterslips evolved on an infinite fault plane with homogeneous VS parameters for infinite time. It should be noted that we use a rate-and-state friction (RSF) law in the numerical simulations only to demonstrate that a physically based afterslip model approaches the asymptotic behavior predicted by the equilibrium-state analysis. The friction law could be replaced by another constitutive formulation, such as CNS friction (Chen et al., 2017) or viscous shear, without altering this asymptotic behavior. In the numerical simulation, we hold the central coseismic region locked and the surrounding afterslip region releases stress through sliding. We understand that realistic faults never have such an idealized setting, but this ideal case helps us to understand the most fundamental principles of the related processes.

### 2.2 Equilibrium State

According to the earthquake cycle theory, a complete cycle should initiate from and ultimately return to an inter-seismic loading state (Scholz, 1998), making the identification of this state necessary. Traditional theory defines afterslips as the aseismic slips following the co-seismic slips and when afterslips drop to background rates, inter-seismic loading starts. Nevertheless, there are two limitations in such a definition. Firstly, inter-seismic loading originates from tectonic loading, which persists during the post-seismic period. i.e., during the post-seismic period, the inter-seismic loading is non-zero. Secondly, it is commonly assumed that the inter-seismic loading rate equals the far-field plate loading velocity. However, fault slip velocity varies with distance from the locked portion, thus the inter-seismic loading rate varies in space as well (Zhao et al., 2022).

In our RSF simulation, a stress loading rate is assumed, which essentially is the inter-seismic loading term and is calibrated from afterslip models. Because tectonic loading continues throughout the post-seismic period, the observed post-seismic slips are simultaneously contributed by both the background and co-seismic loading. As the whole system is linear, the contribution can be viewed as a linear summation of responses of both loading mechanisms. The afterslip is considered as the direct product of co-seismic loading, while the stable tectonic loading should lead to a stable velocity distribution, namely the equilibrium state, which stably releases the loaded tectonic stresses over time. For the inter-seismic period, the equilibrium state directly releases stress over VS patches while loading locked VW patches. In afterslip periods, such state maintains, which is the velocity state to be calibrated from simulated post-seismic slips to recover true afterslips. The algorithm to derive the equilibrium state is stated as follows.

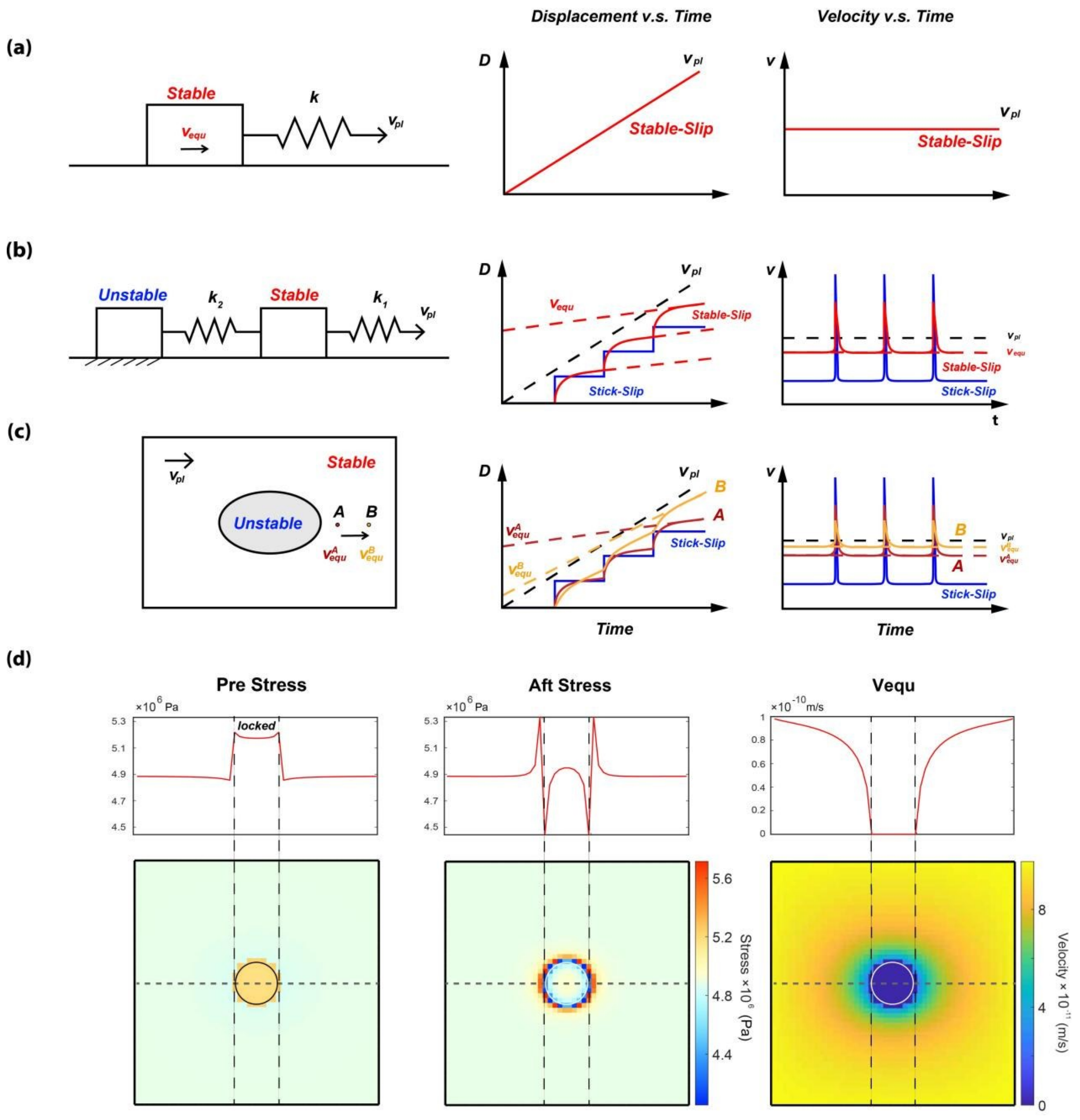

**Figure 2.** Illustration of the equilibrium velocity. (a) A single spring-slider model. For a stable velocity strengthening ($a - b > 0$) slider, the equilibrium velocity $v_{equ}$ is identical to the background loading rate $v_{pl}$. (b) Two-body spring-slider model. The coupled system consists of velocity weakening (unstable) and velocity strengthening (stable) sliders. The kinematic interaction results in a $v_{equ}$ lower than $v_{pl}$. (c) Dual friction (VS and VW) distribution on a 2-D fault plane. The fault plane consists of the co-seismic zone (unstable) and the afterslip zone (stable). A spatial distribution of $\boldsymbol{v_{equ}}$ exists within the afterslip zone, which progressively converges to $v_{pl}$ at greater distances from the co-seismic patch (e.g., at points A and B). (d) Numerical solution of the stress just-before and after the co-seismic rupture and the spatial distribution of $\boldsymbol{v_{equ}}$ on a 2-D fault plane. The figures above show the cross-section indicated by the gray dashed line, with the VW zone delineated by the black dashed lines. Note that panel (d) is obtained from a dynamic earthquake-cycle simulation using the full RSF, which aims to verify that the equilibrium state derived from the ss-RSF analysis is indeed attained in a self-consistent earthquake-cycle simulation.

We use Figure 2 to demonstrate the stress conditions of the equilibrium state. For the simplest, single spring-slider (VS) case, the equilibrium state is achieved when the sliding velocity equals the far-field loading velocity. This equilibrium state is maintained over time and there is no further stress loading or state change following this status (stable). In such a case, it is straightforward to define the equilibrium state as the state in which the slip velocity $v_{equ}$ does not change over time. The definition of equilibrium velocity can be derived by differentiating Equation 1 with respect to time:

$$\dot{\tau} = \sigma(a-b)\frac{\dot{v}}{v} \quad (5)$$

which indicates that when the fault attains a state of stable inter-seismic creeping (i.e., $\dot{v} = 0$), the stress loading rate on the fault must be zero (temporally stable stress). In the single spring-slider model, the stress loading rate can be analytically derived as:

$$\dot{\tau} = k(v_{pl} - v) \quad (6)$$

where $k$ is the spring stiffness and $v_{pl}$ is the background loading rate. Setting $\dot{\tau} = 0$ immediately yields $v_{equ} = v_{pl}$. When defining the state evolution following a sudden stress loading, we combine Equations 5 and 6, which leads to the evolution equation for slip velocity:

$$v = \frac{v_{equ}}{1-(1-\frac{v_{equ}}{v_{ini}})exp(-\frac{t}{t_c})} \quad (7)$$

in which $v_{ini}$ is the initial afterslip velocity calculated by substituting the co-seismic stress loading $\Delta\tau$ into Equation 1, and the characteristic time $t_c = \sigma(a-b)/kv_{equ}$, which is consistent with Perfettini & Avouac (2004). The solution of $v$ follows a typical exponential shape, dropping from $v_{ini}$ to $v_{equ}$ over a characteristic time of $t_c$.

For a chained two-body spring-slider model consisting of a VW slider (locked in the loading period) connected with a VS slider, the same derivation yields a $v_{equ}$ that is lower than $v_{pl}$ (Figure 2b). An analogy to the single spring-slider case involves a balance of three forces (i.e., plate drag, static slider hold and frictional force) achieved by rewriting Equation 6 as:

$$\dot{\tau} = \begin{pmatrix} k_1 \\ k_2 \end{pmatrix} (v_{pl} - v \quad 0 - v) \quad (8)$$

and an equilibrium state of the VS slider is calculated when setting the stress rate as 0, which yields $v_{equ} = k_1 v_{pl}/(k_1 + k_2)$, while new characteristic time $t_c = \sigma(a-b)/k_1 v_{pl}$ following the same form of Equation 7. Although the cycle behavior of the two-body system is more complicated than the single-body system, there still exists an equilibrium state during the stress loading process, in which the VW slider remains locked, while the VS slider slides at a constant velocity at zero shear stress loading rate.

A similar analogy could be extended to a binary fault system. As mentioned in the previous cases, the VS faults creep during the inter-seismic loading period, releasing stress loading placed on them, while loading the VW patches (Figure 2c). Such a system could be simulated by treating each subfault as a plane fault that interacts with other faults. The stress interaction between two portions can be calculated by:

$$\mathbf{G}_{ij}(v_j - v_{pl}) = \dot{\tau}_i \quad (9)$$

where $i, j$ index the subfaults and $\mathbf{G}_{ij}$ represents the Green's function matrix calculated by the Okada solution (Okada, 1992). Here and throughout, bold italic letters denote vectors (e.g., the slip-rate distribution $\boldsymbol{v}$) and bold roman letters denote tensors/matrices (e.g., the Green's function matrix **G**). We assume that the equilibrium state occurs when inter-seismic stress loading on the VS portion is released by its own creep, yielding zero stress loading rate on those faults. Partitioning the fault plane into the afterslip and co-seismic zones, $\mathbf{G}$ is presented as

$$\mathbf{G} = \begin{bmatrix} \mathbf{G}_{aa} & \mathbf{G}_{ac} \\ \mathbf{G}_{ca} & \mathbf{G}_{cc} \end{bmatrix} \tag{10}$$

where $\mathbf{G}_{aa}$ ($N_a \times N_a$; $N_a$ and $N_c$ denote the number of the afterslip and co-seismic subfaults, respectively) uses afterslip subfaults as both sources and receivers (the VS-VS interaction), $\mathbf{G}_{ac}$ ($N_a \times N_c$) uses co-seismic subfaults as sources and afterslip subfaults as receivers (the VW→VS loading), and $\mathbf{G}_{ca}$, $\mathbf{G}_{cc}$ are defined analogously. Thus the velocity distribution of the equilibrium state should satisfy:

$$[\mathbf{G}_{aa} \quad \mathbf{G}_{ac}]\left\{\begin{bmatrix} \boldsymbol{v}_{\boldsymbol{equ}} \\ 0 \end{bmatrix} - v_{pl}\right\} = 0 \tag{11}$$

Equation 11 shows that the stress loading of each subfault is released by its slip. When all subfaults achieve such a state, the stress loading rate is zero and the creeping faults receive no extra tectonic loading, so the sliding velocity is maintained.

Solving Equation 11 yields the distribution of $\boldsymbol{v}_{\boldsymbol{equ}}$ on the fault (i.e., the equilibrium state) as:

$$\boldsymbol{v}_{\boldsymbol{equ}} = \mathbf{G}_{aa}{}^{-1}[\mathbf{G}_{aa} \quad \mathbf{G}_{ac}]v_{pl} \tag{12}$$

Similar to the two-body spring-slider model in Figure 2b, the locked co-seismic zone places a drag on the afterslip zone, leading to a $\boldsymbol{v}_{\boldsymbol{equ}}$ distribution lower than $v_{pl}$. This drag-effect decays with distance from the locked zone (Figure 2c), thus $\boldsymbol{v}_{\boldsymbol{equ}}$ converges to 0 near the locked patch margin and converges to $v_{pl}$ at infinite distance (Figure 2d).

It is noted that the shape of VW and VS patches can be arbitrarily set, thus the form of the $G$ matrix varies, which cannot be solved analytically. However, this situation could be flexibly simulated through a numerical method, i.e., BIM, as discussed above. Relying on the same theoretical foundation (ss-RSF and Okada's formulation), the afterslip simulated in our program consistently converges to the equilibrium state obtained from the prescribed fault parameters and background loading rate (Figure 2c). There are at least three important features of the equilibrium state:

a. $\boldsymbol{v}_{\boldsymbol{equ}}$ is a heterogeneous velocity field over the fault plane. Different from the scalar of $v_{pl}$, $\boldsymbol{v}_{\boldsymbol{equ}}$ is an array of values (a vector) representing the velocity at each point on the fault plane, and we use bold-italic characters to present such values in this paper. In areas far away from the locked patch, $\boldsymbol{v}_{\boldsymbol{equ}} = v_{pl}$, while $\boldsymbol{v}_{\boldsymbol{equ}}$ decreases to zero near the edge of the locked patch.

b. The equilibrium state is constant throughout earthquake cycles, including the post-seismic period. When determining the afterslip, the contribution from equilibrium states should be removed (e.g., in section 2.3) from slip velocities of VS patches, leaving only the contribution of afterslip. This is an important step when evaluating how much co-seismic loading is released by afterslip (Section 2.3), which is also how afterslip is defined throughout this study.

c. $\boldsymbol{v_{equ}}$ is calculated by ensuring a zero stress rate in the VS zone, but it is important to note that stress is non-zero (Figure 2d). Under ss-RSF, the shear stress is determined from $\boldsymbol{v_{equ}}$ through Equation 1, and because stress and velocity are positively correlated in the VS zone, a lower $\boldsymbol{v_{equ}}$ near the locked patch corresponds to a lower shear stress. This reduction of stress adjacent to a strongly coupled patch constitutes a stress shadow, and it offers a possible physical explanation for the elevated locking and reduced creep observed near shallow megathrust asperities (Lindsey et al., 2021). We emphasize, however, that the stress-shadow concept is debated, since such patterns can also reflect frictional or stress heterogeneity; we therefore present it only as a plausible implication of the equilibrium state rather than a definitive interpretation.

### 2.3 Instability of $R_M$

With the equilibrium state established, we can calibrate the post-seismic slips by subtracting $\boldsymbol{v_{equ}}$ from the afterslip model (Section 2.2). The derived afterslip model is then used to examine the moment ratio $R_M$. We emphasize that we aim to determine the relationship between co-seismic and afterslip in a single earthquake cycle, without a new event to stop the last cycle, thus we maintain the VW patch to be locked to infinite time to evaluate the asymptotic behavior of afterslips. If a simulation is made to describe multiple cycles, the remnant stress would be added to the next cycle to influence the next afterslip behavior. In the current setting, the total afterslip distribution will converge to its asymptotic form $\boldsymbol{v_{equ}}$ in an infinite time, e.g. $\lim_{t\to\infty} \boldsymbol{v}(\boldsymbol{x}, t) = \boldsymbol{v_{equ}}(\boldsymbol{x})$. In such a condition, the stress status converges to that predicted by $\boldsymbol{v_{equ}}$. To recover the situation that co-seismic stress loading is fully released, $\boldsymbol{v_{equ}}$ should be removed from the total afterslip model to recover pure afterslips and its final stress status. After removing the equilibrium state, it is clearly revealed that the effect of afterslip is to remove the co-seismic stress loading in the VS portions. We verify this numerically by simulating afterslip on VW patches of different shapes (triangle, ellipse, square and heart) under a uniform stress drop; in each case the afterslip-released stress approaches the co-seismic loading, and the relative residual stress $R = (\tau_{co} - \tau_{aft})/\tau_{co}$ decays to below 0.1% over ~$10^4$ days (Text S1 and Figure S1 in the Supporting Information), leaving negligible remnant stresses. We emphasize that the calibration for the equilibrium state is a key operation to achieve this recognition.

On the other hand, afterslip simulation also shows an expanding pattern, with the initial afterslip located near the co-seismic area and propagating out to surrounding areas, therefore its moment also depends on the available area of VS faults to host afterslips. Since we are interested in the idealized situation without the influence of finite scales of fault planes, we aim to obtain the solution for an unbounded $(r \to \infty)$ fault plane. Intuitively, the co-seismic stress loading only applies to the area surrounding the rupture zone, which decreases to zero in the far-field. Thus when the available afterslip zone expands in scale, its moment also drops to zero and the total afterslip moment should converge to a fixed value rather than increasing without bound. To evaluate this hypothesis, we implement a series of synthetic tests. The fault plane is meshed into subfaults with binary frictional properties: VW subfaults are placed at the center of the fault to host co-seismic rupture, while VS subfaults surround the VW region to host afterslip. The co-seismic stress of a strike-slip circular crack is introduced to initiate afterslip. A set of typical frictional parameters of VS portions are adopted to simulate afterslip evolution. The specific fault parameter settings are provided in Table 1. Regarding the boundary conditions, the modeled fault is embedded in the half-space with its top edge at 500 km depth, so that the traction-free surface has a negligible influence and the solution approaches the full-space limit. At the subfault resolution used here (1 km), our simulation resolves the afterslip stress field throughout the velocity-strengthening region; the steep gradient and the singularity at the rupture edge introduce only a localized discretization error that has a negligible influence on the overall moment. For a co-seismic rupture scale of 10 km, we parameterize the whole fault plane with scales of 50, 100, 200, 500 and 1000 km. The largest fault scale is two orders of magnitude higher than the co-seismic rupture scale, yet we do not find that the afterslip moment converges to a fixed value in such a broad setting (Figure 3b). The numerical test clearly demonstrates that for the same co-seismic rupture loading, the afterslip moment scales with the area of the surrounding VS fault plane, instead of maintaining a fixed ratio to co-seismic moment. Thus, in the numerical simulation $R_M$ is determined by the fault model parameterization instead of being a physical factor linking co-seismic and afterslip. If the real afterslip inversion met with a similar issue, the afterslip moment should scale with the available fault area to host the slips as well as observation time that allows for the afterslips to propagate to the edge of available faults. This mechanism at least partially explains the wide range of $R_M$ reported by Churchill et al. (2022).

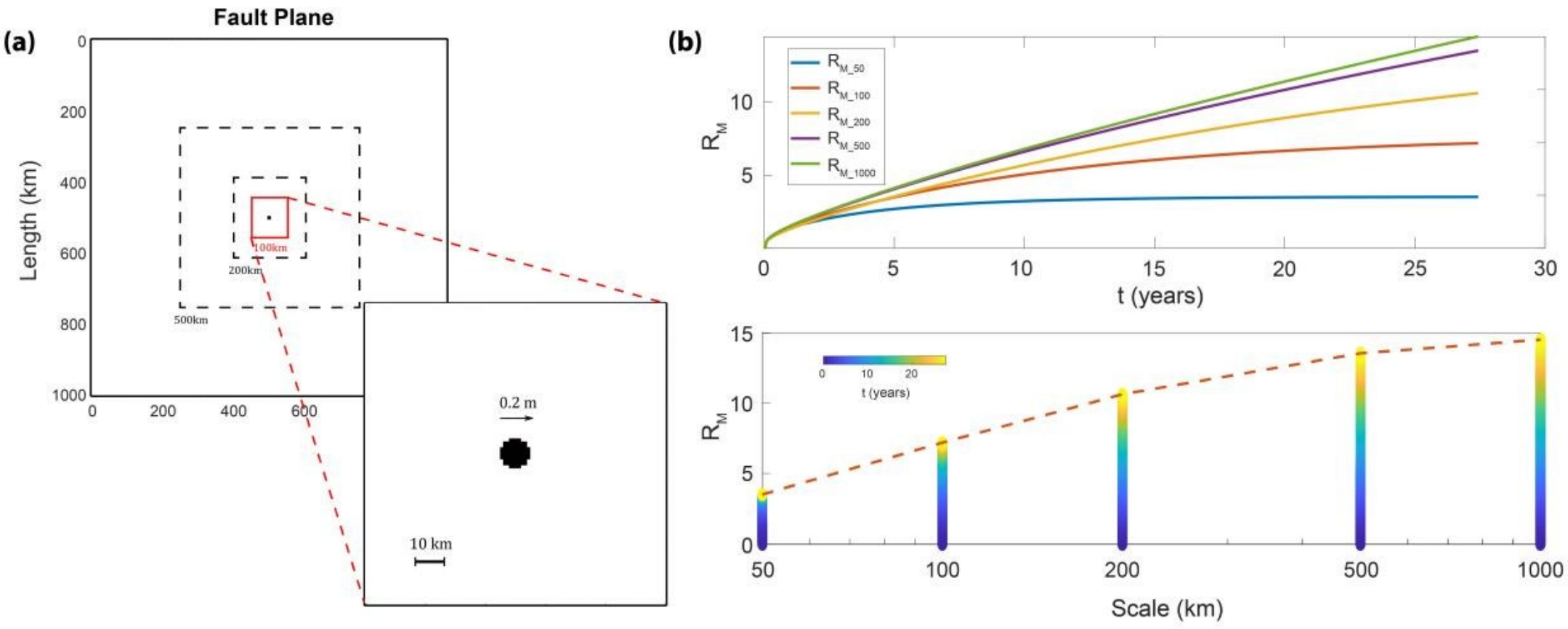

**Figure 3.** Synthetic tests for the convergence behavior of afterslip moments. (a) Fault plane setting with variable afterslip ranges. The co-seismic zone is a circular region with a diameter of 10 km, while the remaining region is the afterslip zone controlled by velocity strengthening friction. (b) Simulation results of the moment ratio between afterslip and co-seismic releases ($R_M$) at different spatiotemporal scales. Red dashed line denotes the temporal maximum $R_M$ as a function of spatial scale, which shows no convergence trend.

**Table 1**

| **Fault Parameters** | **algebraic symbols** | **values** |
|---|---|---|
| Subfault length (along strike) | *dx* | 1 km |
| Subfault width (along dip) | *dy* | 1 km |
| Strike | *strike* | 45° |
| Dip | *dip* | 70° |

| Fault Parameters | algebraic symbols | values |
|---|---|---|
| Depth of the top edge of the fault | *depth* | 500 km |
| Slip type | | *strike slip* |
| Time step | $\Delta t$ | 1 day |
| Shear modulus | $\mu$ | $3.3 \times 10^{10}$ Pa |
| Poisson's ratio | $\nu$ | 0.25 |
| Friction parameter | $\sigma(a-b)$ | $5 \times 10^{4}$ Pa |
| Loading rate | $v_{pl}$ | $2 \times 10^{-10}$ m/s |

The "peculiar" non-convergence behavior of afterslip moments could be easily understood by a thought experiment (Figure 4): neglecting all complicated frictional parameters, when evaluating the final moment release of afterslip, we could consider an infinite fault plane at binary states (Figure 4), i.e., co-seismic slips are hosted by VW patches in the center, which are surrounded by VS creeping sections. Initiated as a uniform co-seismic slip in the center, stresses are loaded in the surrounding area. Afterslip is thus driven by the shear stresses and its effect is to release such stresses. As long as there is residual stress in the VS zone, afterslip continues, until all co-seismic stress is released. This assumption is consistent with the limited remnant stress revealed by numerical simulation (Figure S1). Therefore, the final state of the fault, when afterslip stops, is that all stress loading is released, i.e. zero stress status. Since the slip system is a deterministic problem, the state associated with zero stress represents the final accumulated afterslips. Then the next question is what kind of slip distribution introduces zero stress over the fault. It is known that the shear stress on the fault plane is zero when a uniform slip is applied. Since the VW patch remains locked in the afterslip period, its final slip is uniform as D, when the entire VS region achieves a uniform slip of D, it compensates to the co-seismic slips, achieving a uniform displacement of D over the infinite fault plane, thus the total stress status is zero. Therefore, we conclude that the final accumulated afterslip is uniformly D (identical to the co-seismic slip) over the VS area. In such a case the moment of afterslip is $M_{aft} = \mu DS$, where $\mu$ is the shear modulus and $S$ is the VS area, which is infinite for an infinite fault, thus $M_{aft}$ is infinite (non-convergent). It is also noted that the final state depends on co-seismic slip ($D$) instead of its moment, thus there is no relationship between the co-seismic moment (arbitrary) and afterslip moment (infinite). Therefore, the ratio between afterslip and co-seismic moment will not converge to a finite value, which is consistent with the numerical experiment (Figure 3).

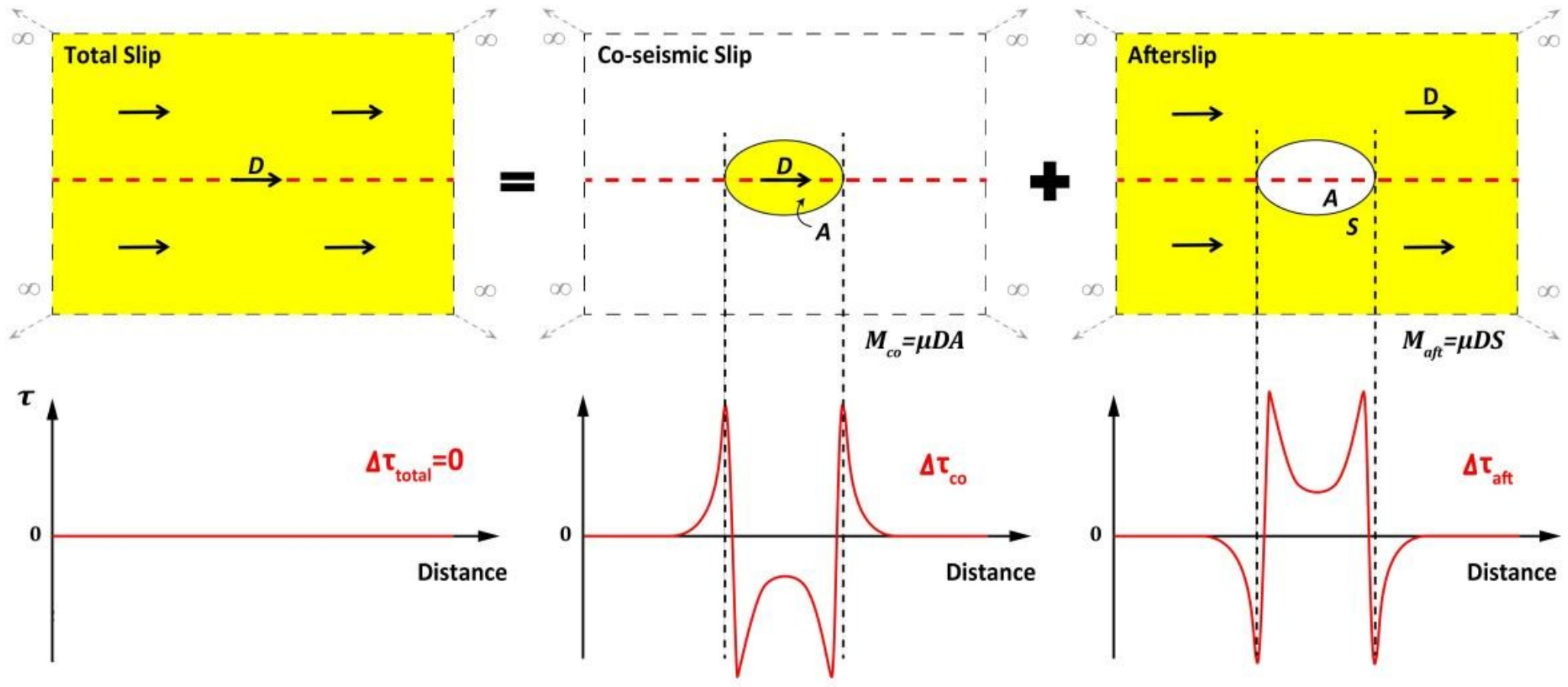

**Figure 4.** A thought experiment to demonstrate the infinite moment release of afterslip. Slip distributions are plotted in the top row and associated stress variations along the red dashed line are plotted in the bottom panel. As a linear system, the full-space uniform slip $D$ is expressed as the sum of co-seismic slip (the patch of area $A$) and surrounding afterslip (the patch of area $S$), with their respective stress changes $\Delta\tau_{co}$ and $\Delta\tau_{aft}$ summing to zero.

Again, in this experiment the final state of afterslips is determined by solving when the uniform stress state is zero. The only assumption here is that afterslip continues when there is remnant shear stress on the fault, without assuming the detailed form of frictional laws. Basically, this condition could be realized by most velocity (shear rate) determined frictional laws, e.g. RSF, CNS or viscous shear band. In summary, under an ideal condition, i.e., the fault area is infinite, afterslip of any co-seismic rupture will have infinite magnitude. In practice, due to the finite dimensions of real faults, the non-convergence of afterslip moment may not occur, while its moment may strongly depend on the fault area allowing for such slips and observation time. This mechanism could be the major reason for the larger variability of $R_M$ in observational studies. Therefore, we conclude that the moment ratio $R_M$ is not a reliable metric for scaling afterslip and should be replaced by an alternative standard.

### 2.4 Conservative Rupture Force

In the above section, we demonstrate that $R_M$ is scale-dependent and thus not a reliable metric; in this section, we seek an alternative conserved quantity that links co-seismic rupture and afterslip. For an infinite planar fault in a homogeneous full-space, Zheng & Rice (1998) showed that the slip-induced stress ($\varphi$) alteration on the fault plane integrates to zero over the entire plane,

$$\iint \varphi(x,z,t)dxdz = 0 \tag{13}$$

for any slip history and distribution. In the quasi-static limit considered here, $\varphi$ reduces to the static stress change. Physically, the result states that a slip distribution on a planar fault, being composed of self-equilibrated double couples, exerts no net force on the fault plane. For the case of stress loading on an infinite planar fault by a rupture in a compact area, equation 13 can be further written as:

$$\iint_{A^+} \varphi(x,z)dA + \iint_{A^-} \varphi(x,z)dA = 0, \qquad thus$$

$$\iint_{A^+} \varphi(x,z)dA = -\iint_{A^-} \varphi(x,z)dA = CRF \tag{14}$$

in which $A^-$ and $A^+$ represent the stress drop (negative) and load (positive) areas, respectively. Equation 14 essentially means that for a distributed rupture, the absolute value of the area integration of stress drop areas should equal that of stress load areas. At the same time, the stress drop area generally coincides with the co-seismic slipped area, while stress load area is the afterslip area, thus equation 14 may provide a link to quantitatively relate co- and afterslips. We further name this common value as the **conservative rupture force** (CRF): "conservative" refers to the equality between the two regions and through two processes, and "force" is related to the dimension (N) of the area-integrated stress. Throughout this paper, the CRF is only computed for the stress component resolved along the slip direction, which is more representative. It is also proved that after calibrating the equilibrium states, the co-seismic loading in the VS region will be eventually released by afterslips at an infinite time, thus CRF also represents the integration of final stress release of afterslips. To summarize the three related stress processes, CRF is preserved in three different spatial and temporal domains.

$$CRF = \iint_{A^+} \Delta\tau_{co} dA = -\iint_{A^-} \Delta\tau_{co} dA = -\iint_{A^-} \Delta\tau_{aft} dA \tag{15}$$

where $\Delta\tau_{co}$ and $\Delta\tau_{aft}$ represent stress changes of co-seismic and afterslip stages respectively, while $A^+$and $A^-$ represent the positive and negative areas of either stress regimes, respectively. A concrete example of CRF is the circular crack with uniform stress drop, for which this equality is derived analytically (Text S2). Real faults, however, are embedded in an elastic half-space bounded by a traction-free surface, so it is natural to ask whether the CRF still holds for the half-space situation. We confirm this result by applying Betti's reciprocity theorem to a uniform-slip test field; the integral of the fault-plane traction vanishes in the half-space as well (Text S3 in the Supporting Information, geometric setup in Figure S3), thus equation 14 also holds up for a planar fault in a half-space without invoking any particular rupture geometry or slip distribution.

To verify the conservation numerically and to delineate its scope of applicability, we perform synthetic tests. We focus on the two factors most likely to affect the result in practice: the depth of the dislocation and the spatial extent of the modeled domain. The spatial extent, in turn, controls the size of the afterslip region surrounding the rupture, and is the direct counterpart of the moment-ratio analysis in Section 2.3, where the afterslip moment failed to converge as the domain grows. We therefore design tests in which the fault depth and the spatial extent are varied independently, while the rupture geometry and slip distribution are maintained as a pure strike-slip circular dislocation with a uniform slip. For each case, we compute the co-seismic stress change on the discretized fault plane and integrate it separately inside and outside the rupture. We denote the integral of the co-seismic stress drop within the rupture by $F_{co}^-$, and the integral of the co-seismic stress increase outside the rupture by $F_{co}^+$. Here $F_{co}^+$ is integrated over the entire modeled domain (so $F_{co}^+/F_{co}^- \rightarrow 1$ as the domain grows to infinity).

We compute the CRF ratio $F_{co}^+/F_{co}^-$ for all cases (Figure 5), and $F_{co}^+/F_{co}^-$ increases from ~85% at 20×20 km² to ~97% at 100×100 km² (Figure 5a) and continues toward ~99% at the largest scale (1000 km; Figure 5b), while varying by less than ~5% across the tested depths. The residual shortfall from 100% reflects the finite extent of the modeled domain: as the scale grows and a larger fraction of the external stress increase is captured, $F_{co}^+/F_{co}^-$ converges toward the theoretical value. These results confirm our theoretical analysis that the CRF also holds in a half-space, independent of the depth. Unlike the seismic moment, this conserved quantity is an intrinsic property of a dislocation rupture. (Supplementary tests over different rupture shapes and slip distributions are presented in Figure S4)

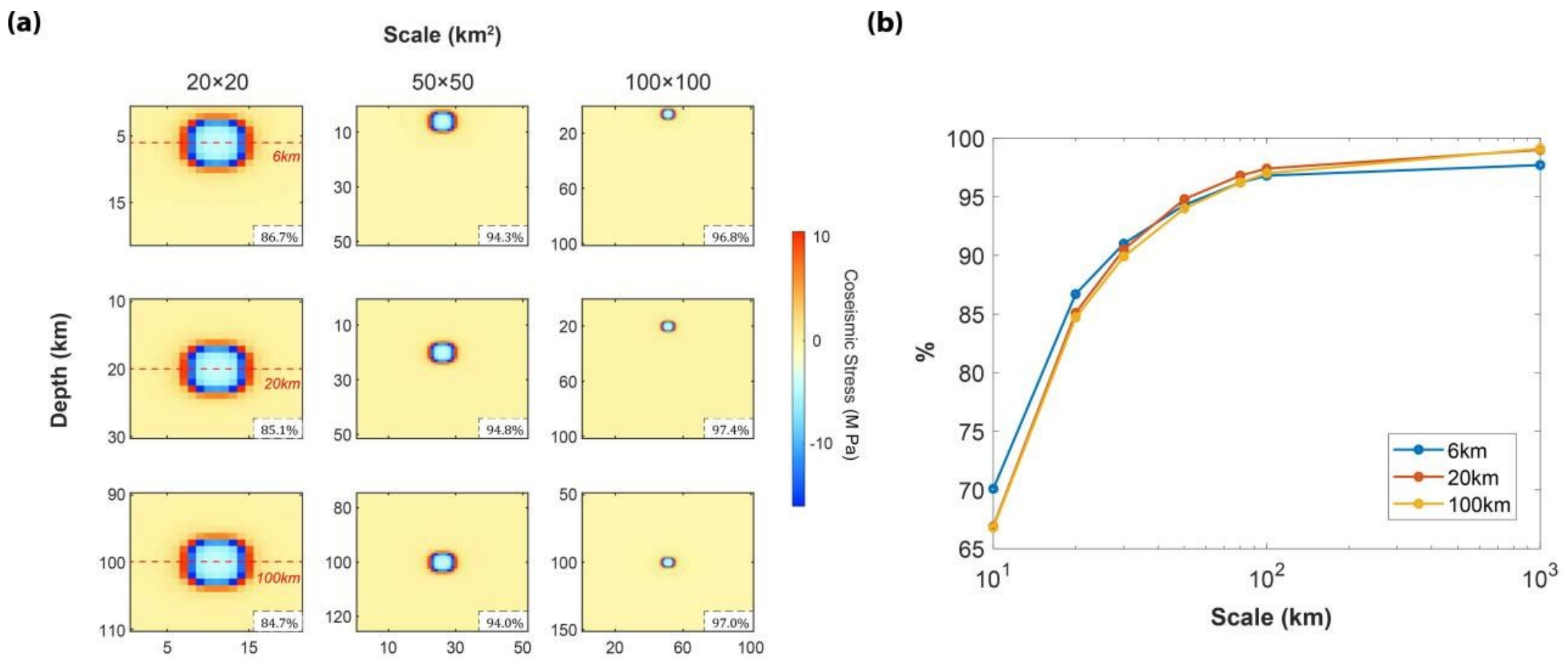

**Figure 5.** Synthetic tests for the conservative rupture force (CRF) under different model scales and rupture depths. The rupture is prescribed as a pure strike-slip circular dislocation of radius 4 km with a uniform slip of 0.8 m. (a) Co-seismic stress distribution on the fault plane for model scales of 20×20, 50×50 and 100×100 km² (columns) and central rupture depths of 6, 20 and 100 km (rows). The central rupture depth is delineated by the dashed red line, and the inset in each panel gives the CRF ratio $F_{co}^{+}/F_{co}^{-}$, which increases from ~85% (20×20 km²) to ~97% (100×100 km²) while differing by less than ~2% among the three depths. (b) $F_{co}^{+}/F_{co}^{-}$ as a function of model scale for the three rupture depths, showing that $F_{co}^{+}/F_{co}^{-}$ converges toward 100% as the model scale grows and is nearly independent of the rupture depth.

## 2.5 Correction for Real Faults

When applying the CRF calculation to a natural afterslip process, several practical issues ought to be considered. The conservation derived in Section 2.4 holds for an idealized planar fault on which the entire non-ruptured region releases the co-seismic stress loading. On a real fault, however, only part of the plane releases stress during the post-seismic period, while the remaining portions could stay locked; the fault of afterslip model always has a finite extent, and the observation covers only a finite time, so the co-seismic loading is only partially released. In addition, the equilibrium state (Section 2.2) ought to be removed from the kinematically inverted afterslip model before the CRF integrals are evaluated. With respect to the stress integral theory described in Section 2.4, we delineate the relevant regions directly from the stress changes implied by the slip models, as described below.

The three forces used in the following are defined from the stress changes calculated by the kinematically inverted slip models. From the co-seismic slip model, we compute the co-seismic stress change $\Delta\tau_{co}(x)$; the negative stress (stress drop) area $A^-$ is defined by the set of subfaults on which $\Delta\tau_{co} < 0$, thus we compute the co-seismic stress drop CRF, defined as:

$$F_{co}^{-} = -\int_{A^-} \Delta\tau_{co}\, dA \tag{16}$$

From the afterslip model we compute the afterslip stress change $\Delta\tau_{aft}(x)$; the afterslip stress-release area $B$ is the set of subfaults on which the afterslip has released stress (i.e., $\Delta\tau_{aft} < 0$),

over which:

$$F_{aft}(t) = -\int_{B} \Delta\tau_{aft}(t)\, dA \tag{17}$$

And the co-seismic stress increase over the afterslip stress-release area is:

$$F_{co}^{+} = \int_{B} \Delta\tau_{co}\, dA \tag{18}$$

By the conservation of Section 2.4, $F_{co}^{-}$ equals the co-seismic stress increase integrated over the entire non-ruptured plane; $F_{co}^{+}$ is therefore the fraction of co-seismic loading that actually falls on regions that are capable of afterslip. Although the complementarity of the co-seismic rupture and afterslip distributions has been a widely adopted premise in afterslip studies (e.g., Hillers et al., 2006; Marone et al., 1991; D'Agostino et al., 2012; Gualandi et al., 2014), our stress-based delineation does not require this assumption, since such a definition is more consistent with the stress interaction theory from which CRF is defined.

Because a real fault has a finite extent and only a portion of it releases stress, the afterslip-released force $F_{aft}(t)$ measured over the finite release area is smaller than the ideal value. We therefore define a dimensionless ratio: $R_F(t) = F_{aft}(t)/F_{co}^{-}$, which measures the fraction of the co-seismic CRF released by afterslip up to time $t$. In the idealized limit in which the entire non-ruptured plane releases the loading ($F_{aft} \rightarrow F_{co}^{-}$), $R_F$ would approach 1. However, in a real afterslip model, $R_F$ saturates at:

$$K_s = \frac{F_{co}^{+}}{F_{co}^{-}} \tag{19}$$

i.e., the proportion of the co-seismic CRF that falls on the stress-release (VS) region. We then define a corrected $R_F$ (namely $R_F^{cor}$) as:

$$R_F^{cor}(t) = \frac{1}{K_s}\frac{F_{aft}(t)}{F_{co}^{-}} = \frac{F_{aft}(t)}{F_{co}^{+}} \tag{20}$$

which measures the release efficiency within the afterslip portion and converges to 1 as $t \rightarrow \infty$. $K_s$ thus plays the role of a correction factor: it equals 1 when the entire non-ruptured region releases stress, and is smaller when a larger part of the fault remains locked. Synthetic tests of $K_s$ performance are plotted in Figure 6, which achieves a discrepancy of about $\pm 3\%$, when the afterslip scale varies within 1-10 times the co-seismic scale. This variation is acceptable when dealing with real data.

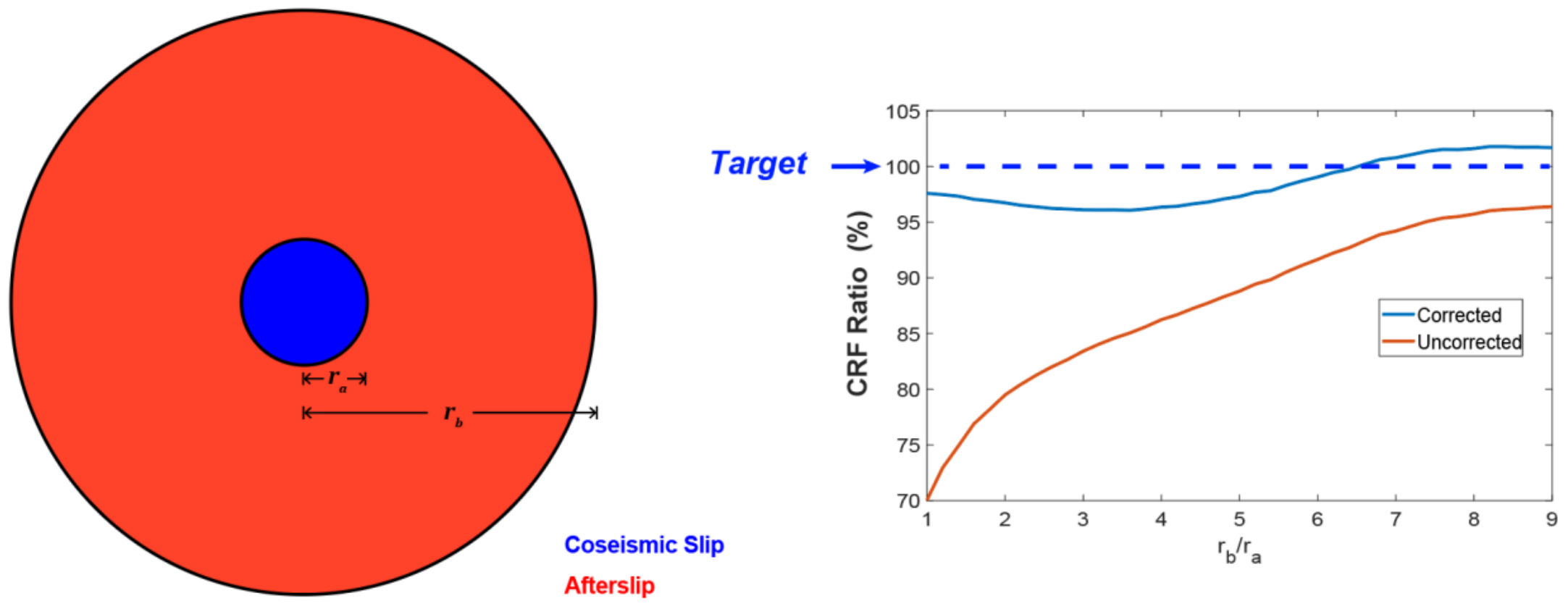

**Figure 6.** Verification of the spatial correction method for 2-D finite fault model. The uncorrected curve represents how the CRF ratio $R_F$ (computed following Section 2.4) varies with the spatial extent of the simulated fault, while the corrected curve shows the $R_F^{cor}$ for an infinite fault plane.

To validate the area integration of afterslip releases and performance of time-dependent CRF ratio, we conduct synthetic tests using different fault parameters on the circular crack model. Given the discretization errors and the finite boundary problem, we apply the spatial correction to the stress released by afterslip within an annular region surrounding the co-seismic zone (red area in Figure 7a), where the discretization errors and the boundary singularities are not significant. The results obtained within the annular region can then be scaled to the entire fault plane using $K_s$. The derived time-dependent CRF ratio $R_F^{cor}(t)$ is plotted in Figure 7b. $R_F^{cor}$ converges to 100% under various frictional parameter settings, and parameter variations only affect the convergence rate. Combining all the theoretical derivations and synthetic tests, we conclude that the CRF is a link to connect co-seismic and afterslip, which provides a more stable and robust metric for scaling afterslip compared to the traditional moment ratio $R_M$.

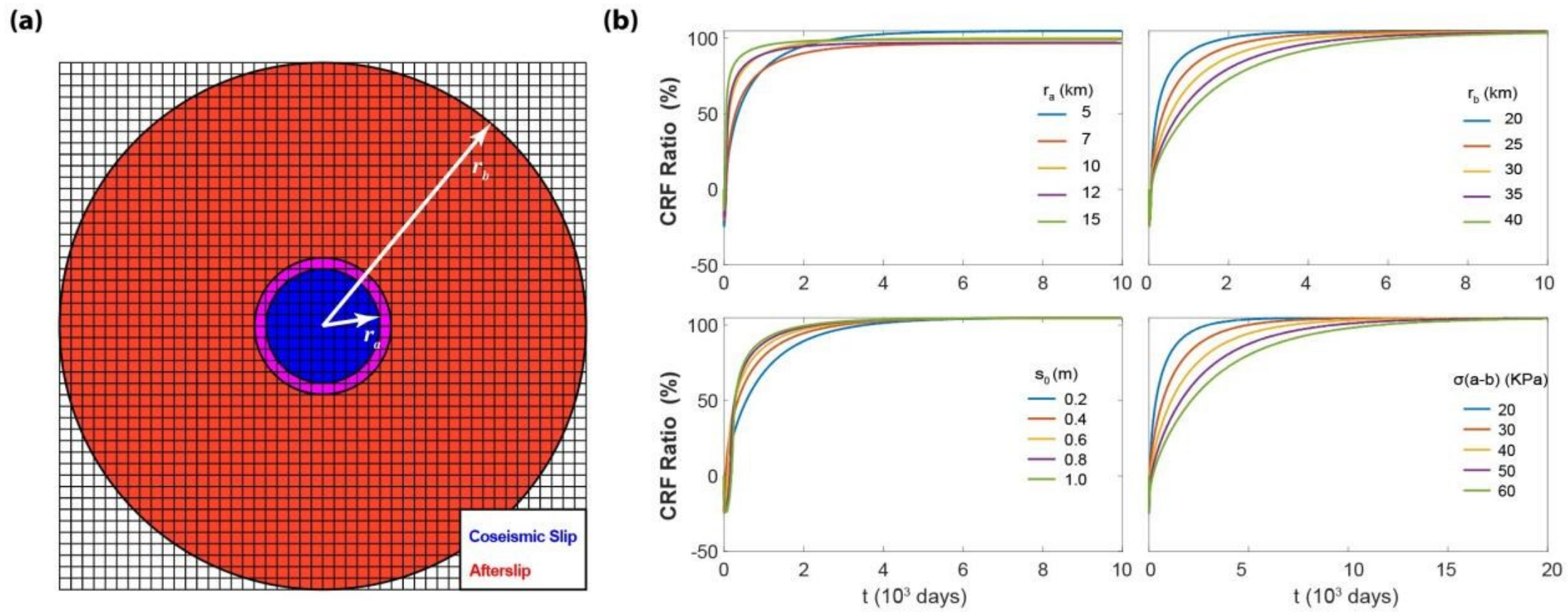

**Figure 7.** Temporal evolution of the corrected CRF ratio. (a) Gridded fault plane. The blue and red regions represent the co-seismic rupture zone with radius of $r_a$ and the afterslip zone with radius of $r_b$, respectively. The magenta ring highlights the area where numerical errors in co-seismic stress calculations are prominent. (b) Temporal evolution of the corrected CRF ratio under different frictional parameters, showing convergence to 100% with a rate that depends on the frictional parameters; $s_0$ denotes the maximum co-seismic slip. The reference group uses $r_a = 5$ km, $r_b = 25$ km, $s_0 = 0.4$ m, $\sigma(a-b) = 2 \times 10^4$ Pa.

# 3 Application, Results & Discussion

## 3.1 Conservative Rupture Force of the Gorkha Earthquake

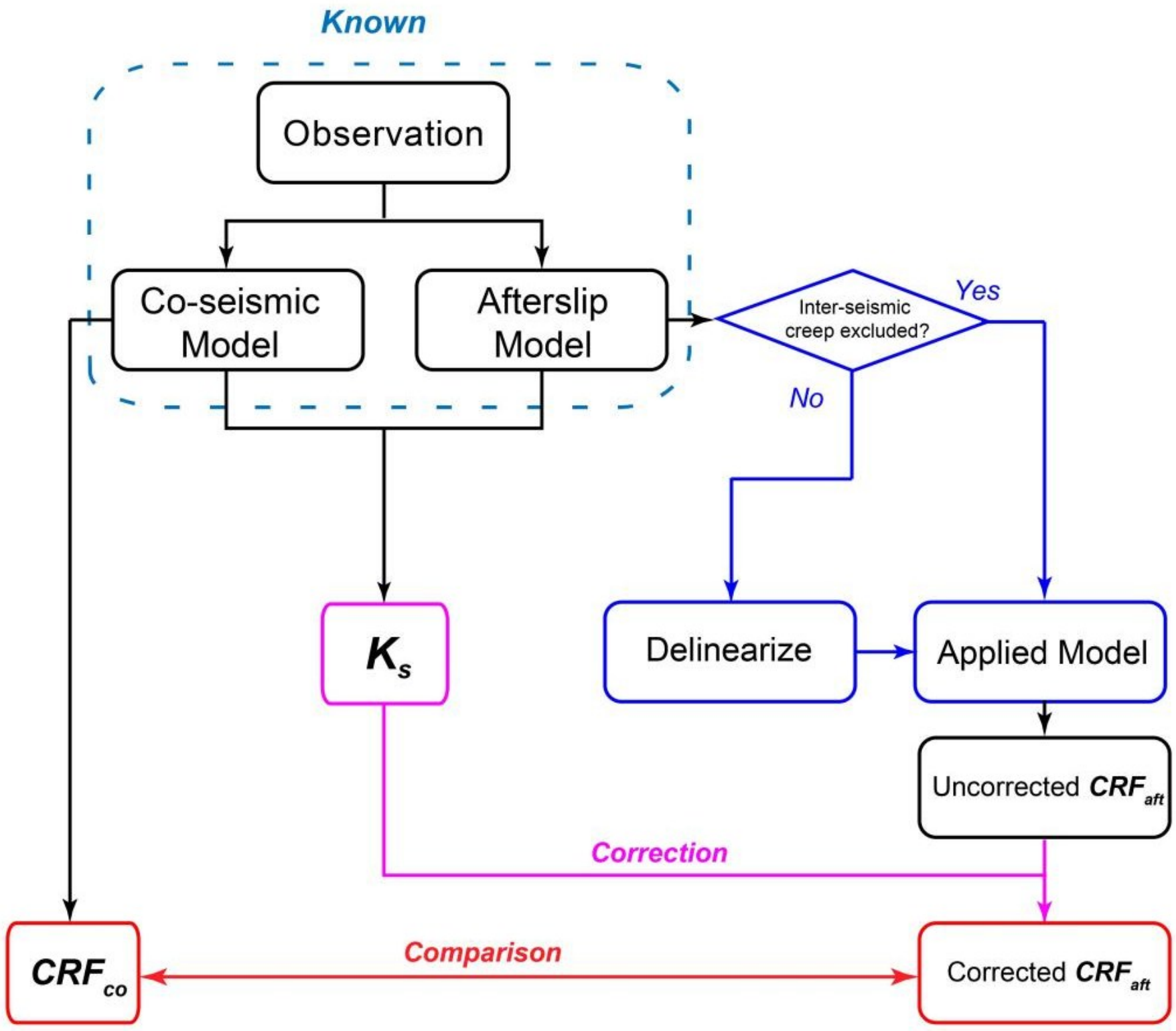

**Figure 8.** Flowchart for the application of CRF in actual earthquake cases.

Figure 8 summarizes the workflow of calculating $R_F$ for a real earthquake, which comprises four steps.

(i) Model construction: the co-seismic and afterslip slip models are derived from geodetic observations (e.g., GNSS and InSAR), from which the stress changes are computed.

(ii) Removal of the long-term inter-seismic loading signal: because the inter-seismic loading persists throughout the post-seismic period, it must be excluded before the afterslip is quantified. If the afterslip model was derived without removing this component, the long-term linear trend representing the inter-seismic velocity is removed from the observations (delinearize). The $\boldsymbol{v}_{equ}$ distribution could be obtained by solving equation 12 with a given VS and VW pattern. On the other hand, if the background linear trend of GPS data has already been excluded before the afterslip model inversion (e.g., Malservisi et al., 2015; Wang & Fialko, 2018), this is equivalent to removing the inter-seismic related signal from data; thus the afterslip model is considered to have $\boldsymbol{v}_{equ}$ removed, and is used directly. The resulting model is the "applied model".

(iii) Uncorrected CRF: the afterslip stress-release area and the uncorrected afterslip force,

$F_{aft}$, are computed from the applied model.

(iv) Corrected CRF: the co-seismic model is used to compute $K_s$. The corrected CRF is then obtained directly as $F_{aft}/K_s$, and its ratio to the co-seismic CRF $F_{co}^{-}$ gives the corrected ratio $R_F^{cor}(t) = F_{aft}(t)/F_{co}^{+}$, the time evolution of which quantifies how much of the co-seismic loading on the releasing portion of the fault has been released by afterslip.

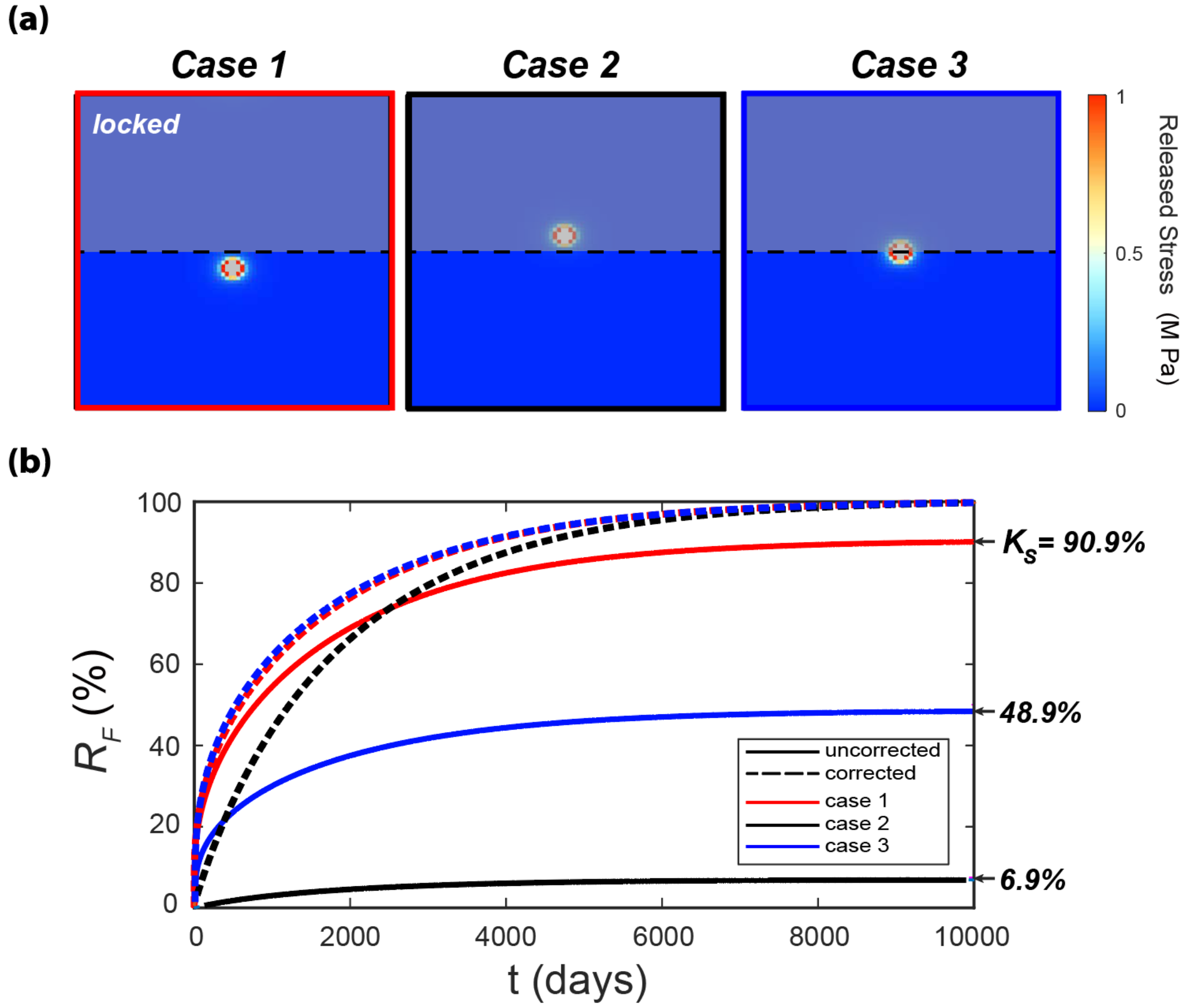

**Figure 9.** Theoretical test of the spatial correction for different fault-zone configurations. (a) Fault-plane geometry for three cases: the fault is divided into a locked region (upper half) and a creeping region (lower half), and a co-seismic rupture is located in the creeping region (Case 1), locked region (Case 2), or straddling the locked/creeping boundary (Case 3). (b) Time evolution of the CRF ratio $R_F$ (solid: uncorrected, dashed: corrected). The uncorrected curves saturate at the correction factor $K_s$, which equals 0.91, 0.07, and 0.49 for Cases 1-3 (the proportion of the co-seismic CRF located in the creeping region), respectively, whereas the corrected curves all converge to ~100%. The spatial correction therefore removes the bias introduced by the locked region and recovers the full release efficiency of afterslip.

To illustrate the meaning of the spatial correction $K_s$ introduced in Section 2.5, we test the workflow of Figure 8 on three idealized fault-zone configurations (Figure 9). In each case the fault is vertically divided into a locked (VW) region and a creeping (VS) region, and the co-seismic rupture is placed in the creeping region (Case 1), locked region (Case 2), or straddling the locked/creeping boundary (Case 3). Because the afterslip releases stress only within the creeping region, the uncorrected ratio saturates directly at $K_s$, the proportion of the co-seismic CRF located in the creeping region, which takes the values 0.91, 0.07, and 0.49 for the three cases, a smaller $K_s$ corresponding to a smaller part of the fault hosting afterslip. After the correction, all three $R_F^{cor}$ curves converge to ~100%, demonstrating that the spatially corrected ratio measures the release efficiency within the slipping portion of the fault, free of the influence from the locked region. This test shows that both uncorrected $R_F$ and corrected $R_F^{cor}$ have physical meaning. The $R_F$ reflects how much of the total loaded forces are released by the afterslip model, and the remnant force could either be released in future afterslip or remain locked to be released by future earthquakes. On the other hand, the $R_F^{cor}$ only reflects how much of the CRFs in the creeping section are released by the afterslip model.

We use the 2015 Gorkha earthquake and its afterslip process as an example to demonstrate the performance of CRF ratio. This event is particularly suitable for such an analysis because its fault geometry was well-constrained and the post-seismic deformation was extensively recorded (Elliott et al., 2016; Liu-Zeng et al., 2020; Wang & Fialko, 2018). Crucially, it represents an almost purely thrust-faulting earthquake, allowing us to restrict our analysis to the corresponding thrust-slip and stress components. Our analysis uses the kinematically inverted afterslip model of Zhao et al. (2025) (Figures 10a,b), which was derived via Full Time-series Inversion (FTI, Yue et al., 2021, 2022, 2026) of post-seismic geodetic data: 54 GNSS stations plus Sentinel-1 InSAR time-series (Wang & Fialko, 2018), with the viscoelastic contribution removed using a regional 3-D viscoelastic model (Zhang et al., 2023). In addition, the afterslip model had also removed inter-seismic slip. FTI is a quasi-static linear inversion that solves for a slip distribution whose temporal evolution follows a logarithmic decay with one or more characteristic times. The preferred afterslip model superposes a fast afterslip (~1 day decay) at the downdip edge of the rupture and a slower one (~100 day decay) further downdip, releasing ~8% of the mainshock moment (Mw 7.12) in 5.5 years. The kinematically inverted co-seismic slip model is from Elliott et al. (2016). Based on geodetic observations indicating that the Main Himalayan Thrust (MHT) remains locked at shallow depths (Li et al., 2018; Stevens & Avouac, 2015) after the 2015 event, we treat all subfaults at depths shallower than 13 km as locked, while the remaining region is treated as the afterslip zone (Figure 10a). The distribution of equilibrium velocity is therefore derived based on the co-seismic/afterslip separation (Figure 10c).

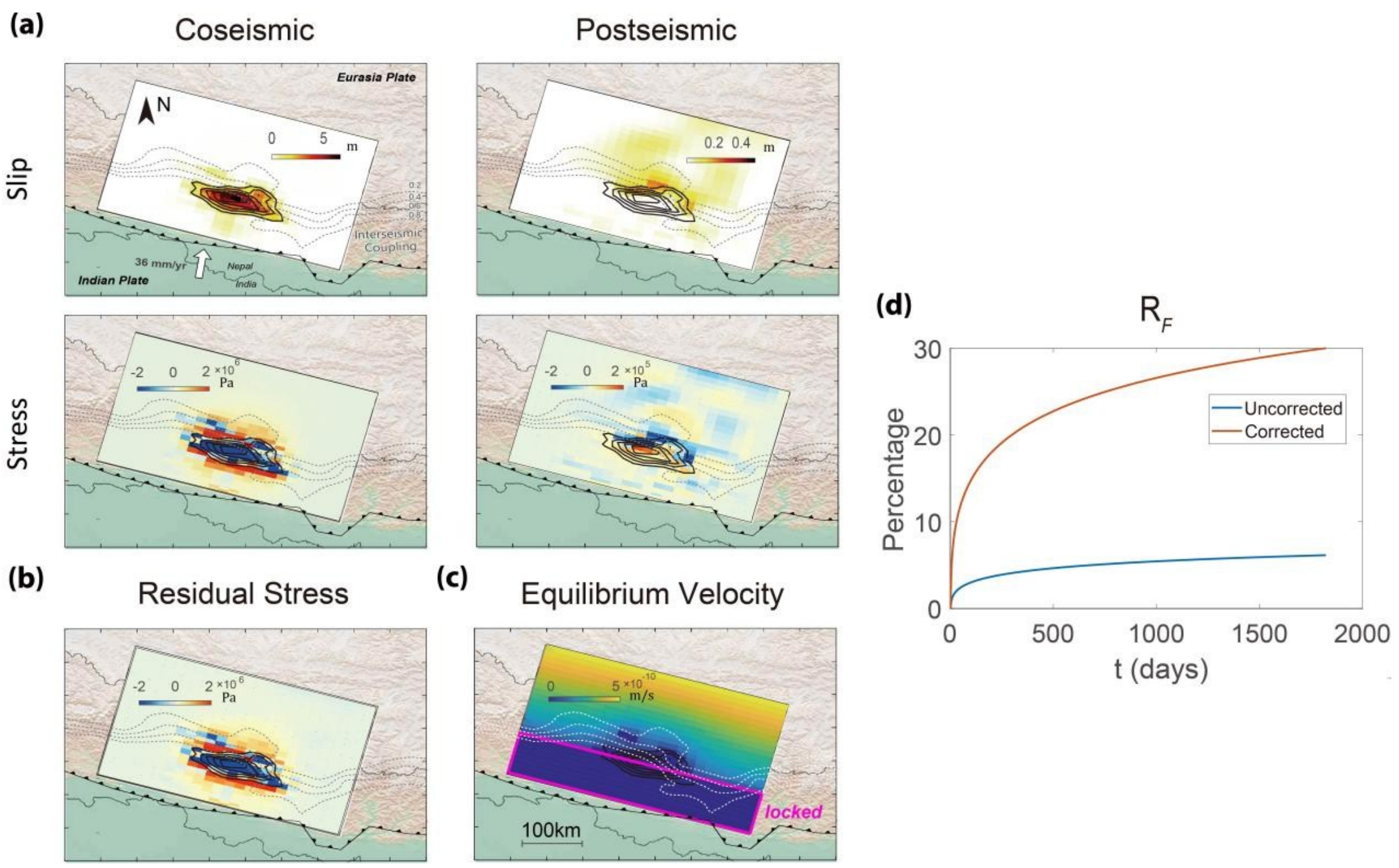

**Figure 10.** Test for the kinematically inverted afterslip model of the 2015 Mw 7.8 Gorkha earthquake. (a) Slip and stress distributions associated with the co-seismic (Elliott et al., 2016) and afterslip (Zhao et al., 2025), respectively. (b) Residual stress (co-seismic stress that has not been released) after 5.5 years of recovery. (c) Distribution of $\boldsymbol{v_{equ}}$ on the fault plane. The magenta line and label outline the prescribed locked region, where $\boldsymbol{v_{equ}}$ is set to zero. In (a)-(c), black contours represent co-seismic slip distribution, and the fault-plane models are overlaid on a regional map of the Gorkha area, which shows the shaded-relief topography, the Indian and Eurasian plates, the surface trace of the Himalayan front (black line with triangles), the interseismic coupling contours (thick grey dashed lines, 0.2-0.8), and the ~36 mm/yr India-Eurasia convergence (white arrow). (d) Temporal evolution of the CRF ratio $R_F$ before and after spatial correction is plotted in blue and orange colors, respectively.

Subsequently, we apply spatial corrections to the kinematically inverted afterslip model. Following the correction, the final result obtained is shown in Figure 10d. Prior to correction, the CRF reaches merely 5% of that loaded by co-seismic rupture 5.5 years after the mainshock. $R_F^{cor}$ reaches 30% after spatial correction. This sixfold difference is mainly attributed to the fact that the up-dip part of the fault is locked after the mainshock. This particular setting is different from the presumed scenario in which the entire non-ruptured area is subject to VS status and hosts afterslip. Other afterslip processes, where both up-dip and down-dip faults host afterslips, e.g., the 2017 Sarpol-e Zahab earthquake (Wang & Bürgmann, 2020; Yue et al., 2022), may yield a smaller correction ratio. The value of $K_S$ itself reflects the locking degree of faults in the post-seismic period. A higher $K_S$ reflects that a larger part of faults hosts afterslips. In summary, the uncorrected $R_F$ (~5%) quantifies the release over the entire fault, whereas $R_F^{cor}$ (~30%) measures the release efficiency within the slipping portion alone.

## 3.2 Possible Explanations for the Deviation

Although the $R_F^{cor}$ of 30% obtained for the Gorkha earthquake falls well short of the theoretical expectation of ~100% established in Section 2.4, this deviation is expected. Several factors likely contribute to this deviation. Most significantly, the influence of other post-seismic mechanisms is not accounted for in our analysis. Unlike the negligible contribution from poroelastic rebound, viscoelastic relaxation played a major role in the post-seismic deformation following the Gorkha earthquake (Tian et al., 2020; Diao et al., 2021). The kinematically inverted afterslip model applied in this study removed this component, meaning that the portion of co-seismic stress released through viscoelastic relaxation is missing from our estimate of the CRF. Similar to afterslip, viscous relaxation is also loaded by co-seismic slip, which is released through ductile deformation in a viscous layer. Zhao et al. (2025) found that the contribution from viscoelastic relaxation can be 1-2 times that of afterslip in terms of moment release. Therefore, how to include its contribution in terms of CRF is worth a focused investigation in future studies.

Other possible factors may also introduce deviations, though their effects are likely minor. For instance, one factor is the complex 3-D geometry of the fault. Our validation of the CRF conservation is based on a synthetic planar fault model, whereas the Gorkha earthquake ruptured with a recognized ramp-flat-ramp-flat geometry (Duputel et al., 2016; Elliott et al., 2016). In such a 3-D setting, the integrated CRF is no longer a simple in-plane vector, which may introduce deviation. Another factor is the heterogeneity of fault properties. In our calculations, all parameters in the VS region are assumed uniform. The presence of asperities or other variations, which are not accounted for, could also lead to deviation in the estimated force.

Although not reflected in this study, the CRF carries additional limitations that may become significant when applied to real faults. One such limitation concerns the uncertain pre-seismic stress state of real faults. The conservation of CRF is strictly valid only for an idealized seismic cycle in which the fault initiates from the stable inter-seismic creeping condition. However, on real faults, when an earthquake occurs, the fault may not have fully released the stress accumulated from the previous seismic cycle or from loading by seismic activities on adjacent faults. In such cases, the CRF released through afterslip could substantially exceed that released co-seismically.

Other potential limitations include, though are not limited to: (1) overlap between co-seismic and afterslip zones due to complex frictional behavior; (2) mutual stress loading among multiple faults involved in the same seismic event; and (3) the absence of a well-defined equilibrium state on faults that host frequent or migrating slow-slip events (SSEs) or persistent locked/creeping segmentation, where the fault may not return to a stable inter-seismic creeping state before the next earthquake (e.g., Bürgmann, 2018). These complications may all challenge the applicability of the CRF framework.

### 3.3 Seismic moment versus CRF

The CRF offers a fresh view to validate the scale of co-seismic and afterslip models, which merits a comparison with the traditional view. For a square co-seismic pattern with dimension $a$ and stress drop $\Delta\tau$, its CRF is approximated as $CRF = \Delta\tau \times a^2$. For the same slip, its seismic moment is defined as $M = \mu \times d \times a^2$. Since its strain drop could be approximated as $\Delta\varepsilon \sim \frac{d}{a}$ while stress drop is estimated as $\Delta\tau \sim \mu \times \Delta\varepsilon$. Its seismic moment could also be approximated as $M = \Delta\tau \times a^3$.

It is clear that CRF and seismic moment have similar mathematical forms, yet a different order of power with characteristic rupture dimension $a$: CRF scales as $a^2$, whereas seismic moment scales as $a^3$. If a physical meaning or practical usage is not considered, we could define a scalar at any order of power of $a$. However, the seismic moment and CRF are incorporated at different conditions with practical purposes. Seismic moment has traditionally been the preferred measure of earthquake size because it governs static co-seismic deformation and low-frequency seismic-wave amplitudes and can be robustly constrained from seismic and geodetic observations; while CRF is derived from the pattern of stress accumulation and release over the seismic cycle. We therefore emphasize that CRF is the physically relevant quantity for problems governed by seismic-cycle stress loading and release. Seismic moment, meanwhile, remains the appropriate parameter for quantifying earthquake size and, with additional assumptions, for radiated energy. Assuming approximately magnitude-independent stress drops of the same order—typically 1-10 MPa (Kanamori & Anderson, 1975)—CRF is proportional to the co-seismic rupture area, whereas seismic moment is proportional to the cube of the rupture dimension under self-similar scaling (for square shaped rupture).

Earthquake moment magnitude is currently defined from its seismic moment, and similarly a force magnitude ($M_F$) could also be defined through its CRF with appropriate calibration to match the moment magnitude. Theoretically, a calibration could be made assuming a square or circular rupture with a constant stress drop. Once such a force magnitude is defined, force magnitudes could be calculated for any earthquake with a finite fault model. The discrepancy between moment $M_F$ and $M_w$ may provide some information about how much the rupture deviates from a reference square/circular rupture, e.g. aspect ratio or slip complexity. The meaning and impact of such a magnitude definition is worth a thorough discussion in future studies.

### 3.4 A Perspective on Earthquake Cycles from Equilibrium State

In this study, we derive the equilibrium state with the purpose of realizing conservative integration for CRFs. At the same time, the equilibrium state also provides a new perspective on the stress status of earthquake cycles. We assume a simple binary fault system, which is composed of a VW zone in the center and a surrounding VS zone. Its associated stress exchange during afterslip and inter-seismic periods is discussed in sections 2.1 and 2.2, respectively, and further summarized here. The key assumption of the binary fault system is that either frictional portion has its characteristic release behavior. The slip behavior of the VW patch is simplified as uniform slip during an earthquake. On the other hand, the VS patch continuously releases its stress loading. Since the VS patch is subject to two types of stress loading, it presents two associated releasing behaviors. Firstly, the VS patch is continuously loaded by a constant tectonic stress, which is released through stable sliding to maintain its state. Therefore an equilibrium sliding state ($\boldsymbol{v_{equ}}$) is derived by solving an equation to ensure the net stress loading rate is zero, i.e., tectonic loading is released through stable sliding in the inter-seismic period (Figure 11). Secondly, the VS patch is occasionally loaded by co-seismic ruptures, and afterslips are thus initiated to release co-seismic stress loading, which is the key mechanism to derive the CRF. Since the fault plane is a linear system, which satisfies the input/output linear summation operation, when the VS patch is subject to two loading mechanisms, its response is also a combination of two outputs. This is the reason why $\boldsymbol{v_{equ}}$ ought to be removed from the observed slow slip processes during the post-seismic period to recover pure afterslips. The temporal releasing history of the VS patch relies on the frictional properties, but the final stage, e.g., CRF and $\boldsymbol{v_{equ}}$, only relies on the configuration (spatial distribution) of VS and VW patches, which also simplifies the calculation of the two mechanisms.

The introduction of a binary fault system also refreshes our perspectives of an earthquake cycle. Traditionally, the inter-seismic, co-seismic, and afterslip periods are considered as three successive processes (Figure 11), which sequentially load and release stress of the fault system. Particularly, inter-seismic loading is considered to follow afterslip when the majority of co-seismic stresses are released. However, in this view, the point when the inter-seismic period begins lacks a clear definition. In the view of a binary fault system, the equilibrium state of $\boldsymbol{v_{equ}}$ is the response of continuous tectonic loading, thus the equilibrium state remains constant through earthquake cycles (Figure 11). On the other hand, afterslip is a process that follows the co-seismic rupture, which lasts for a fairly long time (maybe longer than earthquake intervals) to release its stress loading. In this perspective, inter-seismic and afterslip are two parallel mechanisms that simultaneously act on the VS faults, instead of successive processes. The traditional view sums three sequential slips to construct a uniform slip through an earthquake cycle (Figure 11), while the binary fault view summarizes the co-seismic and parallel inter- and afterslips to construct a uniform slip distribution. Though these two views are equivalent in terms of slip budget, the binary fault view offers a clearer link between (1) co-seismic and afterslip as well as (2) between tectonic loading and equilibrium state.

It is also noted that afterslip decays with time and in the late stage of an earthquake cycle, the velocity distribution on a VS fault should approximate the equilibrium state. However, in real seismic cycles, the interval between successive earthquakes may be too short for afterslip to decay to a level that can be safely disregarded; consequently, the observed inter-seismic creep velocity should be higher than the theoretical equilibrium value. The validation of using the late-stage creeping velocity to represent $\boldsymbol{v_{equ}}$ depends on the earthquake interval and frictional releasing properties, which remains to be validated by more observations.

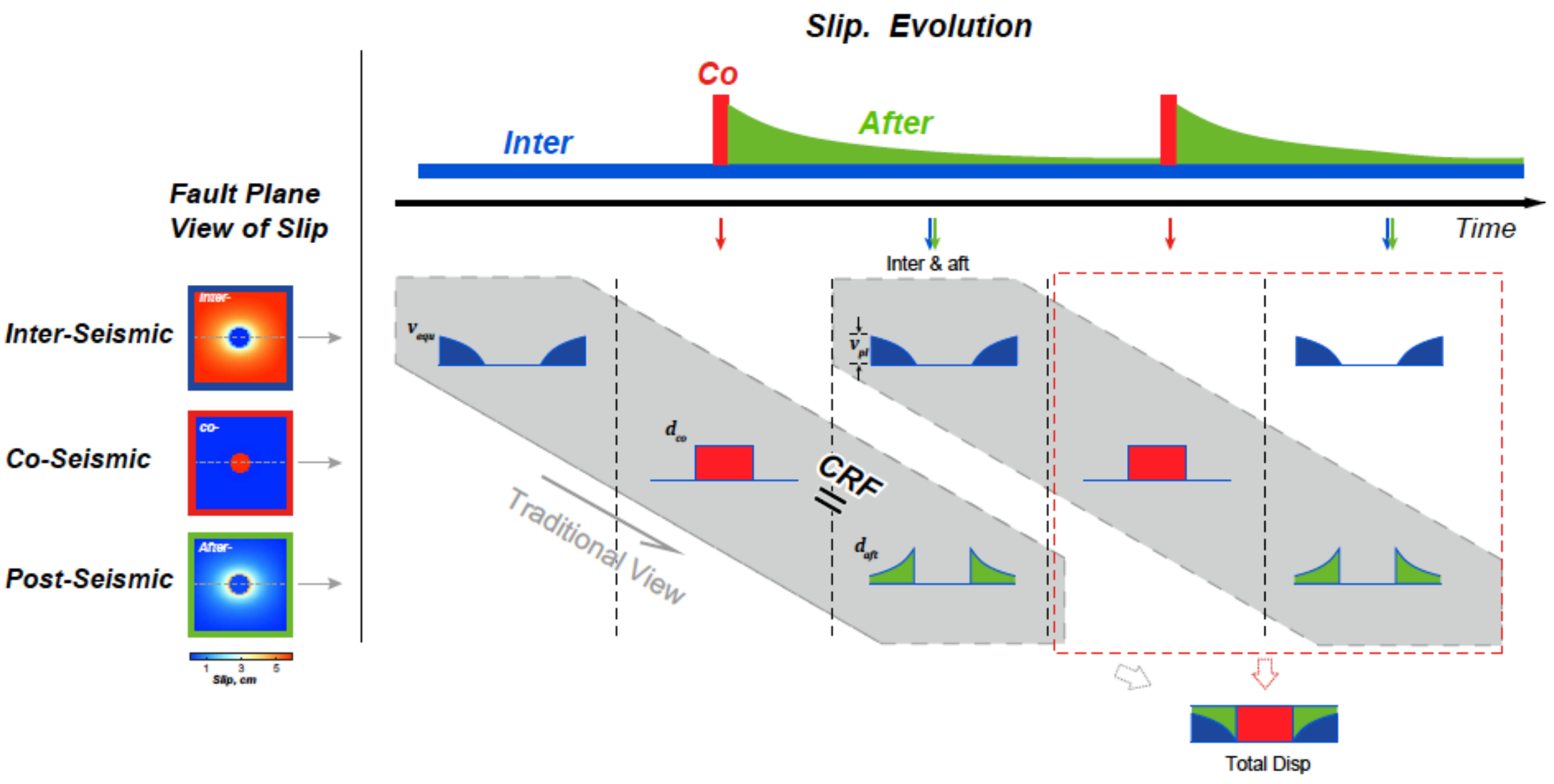

**Figure 11.** Conceptual model of slip evolution on a binary fault. The left column displays the cumulative slip distributions on the fault plane for the inter-seismic (top), co-seismic (middle), and post-seismic (bottom) periods. The top-right timeline illustrates the temporal evolution of these phases, using consistent color coding: blue (inter-seismic), red (co-seismic), and green (post-seismic). Below the timeline, the corresponding slip

profiles along the fault's central axis are shown. The "Traditional View" (grey diagonal) depicts the sequential accumulation of slip over time. The red dashed box highlights the "Parallel View," integrating inter-seismic and afterslip components. The bottom-most profile represents the Total Displacement over a complete earthquake cycle, calculated as the summation of the three distinct periods.

## 4 Conclusions

In this study, we introduce an inter-seismic equilibrium state from which the pure afterslip is isolated. Using the isolated afterslip, we demonstrate that the traditional moment ratio ($R_M$) is an unreliable scaling metric, because its value is controlled by the assumed model scale rather than by any intrinsic physical relationship. To overcome this limitation, we propose the "conservative rupture force" (CRF), defined as the area integral of stress change, and confirm that it is conserved between the co-seismic stress loading and the afterslip stress release through theoretical derivation and numerical tests.

Finally, we apply this framework to the 2015 Gorkha earthquake. With the spatial correction for the finite fault dimensions, we evaluate the time-dependent CRF ratio, which reaches ~30% within the 5.5 years following the mainshock. Although factors like contributions from viscoelastic relaxation introduce some deviations, the time-dependent CRF ratio effectively reflects the real-world release rate of afterslip and provides a reliable measure of the fault's post-seismic locking degree.

## Conflict of Interest Statement

The authors have no conflicts of interest to disclose.

## Acknowledgments

This project is supported by National Science Foundation of China (42525402). The computation and plotting are done in Matlab.

## Data Availability Statement

The kinematically inverted afterslip model of the Gorkha earthquake is available at FigShare (Zhao et al., 2024).

Supporting Information for

# Conservative Rupture Force Relating Co-seismic Rupture and Afterslip Process

Yuxuan Hou[1], Zeyan Zhao[2], Han Yue*[1]

[1]State Key Laboratory of Earthquake Dynamics and Forecasting, SESS, Peking University, Beijing 100871, China, [2]State Key Laboratory of Earthquake Dynamics and Forecasting, Institute of Geology, China Earthquake Administration, Beijing 100029, China

## Contents of this file

This supporting information provides (1) the numerical test that the co-seismic stress loading is fully released by afterslip (Text S1, Figure S1); (2) the derivation of the CRF for a circular crack with uniform stress drop, as an example verifying the Zheng & Rice (1998) conservation (Text S2, Figure S2); (3) the proof that the CRF is conserved on a planar fault in an elastic half-space (Text S3, Figure S3); and (4) the numerical verification of the CRF conservation for ruptures of different shapes and slip distributions (Figure S4).

**Text S1**

We verify the stress-exchange assumption numerically by simulating afterslip on velocity-weakening patches of four shapes (triangle, ellipse, square and heart) under a uniform co-seismic stress drop. In each simulation, the resulting stress loading drives afterslip in the surrounding velocity-strengthening patches, governed by ss-RSF (Figure S1). We quantify the releasing efficiency by the time-dependent relative residual stress $R(\boldsymbol{x},t) = (\tau_{co}(\boldsymbol{x}) - \tau_{aft}(\boldsymbol{x},t))/\tau_{co}(\boldsymbol{x})$ , where $\tau_{co}(\boldsymbol{x})$ is the co-seismic stress loading evaluated immediately after the rupture ($t = 0^{+}$, before any afterslip) relative to the inter-seismic equilibrium state (Section 2.2), and $\tau_{aft}(\boldsymbol{x},t)$ is the stress released by afterslip at time $t$. As shown in Figure S1 (right), $R(\boldsymbol{x},t)$ decays toward zero on each subfault over a long enough time ($\sim 10^{4}$ days), confirming that afterslip eventually releases the entire co-seismic stress loading.

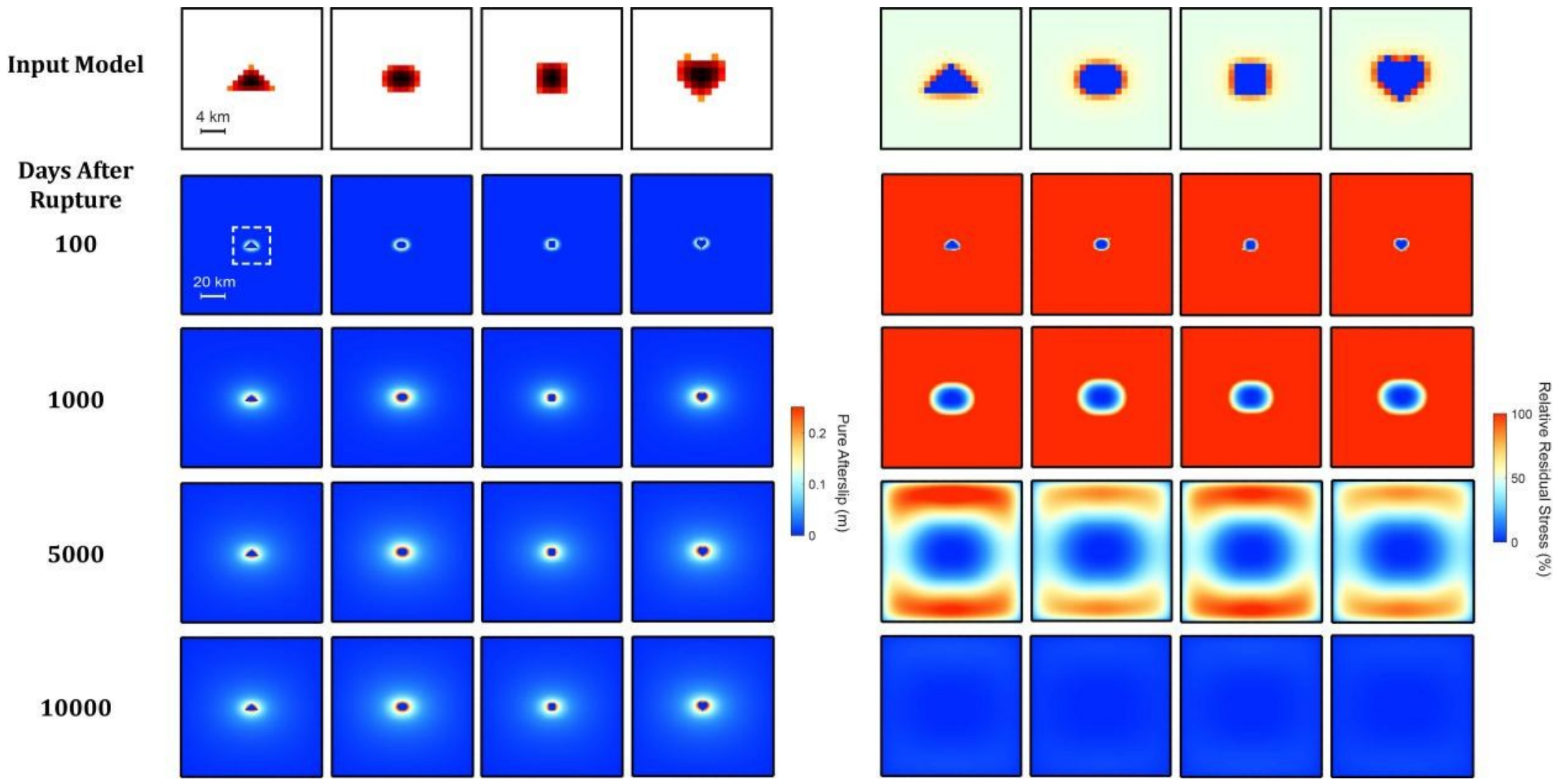

**Figure S1.** Spatiotemporal evolution of pure afterslip (left) and relative residual stress (right) following the co-seismic rupture. The top row illustrates the initial co-seismic slip and stress distributions for the four input models. Subsequent rows show the evolution after rupture (plotted only for the area outside the rupture zone).

## Text S2

As a concrete example of the full-space conservation result of Zheng & Rice (1998) (Section 2.4), we derive the CRF for a circular crack with uniform stress drop. Following Madariaga et al. (2016), the slip distribution required to generate a uniform stress drop along x-direction is given by:

$$u_x(r) = \frac{24}{7\pi}\frac{\Delta\tau}{\mu}\sqrt{{r_a}^2 - r^2} = s_0\sqrt{1-(\frac{r}{r_a})^2} \tag{S1}$$

where $\Delta\tau$ is the stress drop, $\mu$ is the shear modulus, $r_a$ is the rupture radius, and $s_0 = \frac{24}{7\pi}\frac{\Delta\tau}{\mu}r_a$ represents the slip at the center of the rupture.

For a stress-drop distribution $k(\rho)$ inside the slipped zone, the stress change along the slip direction outside the co-seismic zone is derived analytically (Guidera et al., 1975):

$$\tau_{zx}(r,\theta) = -\frac{2}{\pi\sqrt{r^2 - {r_a}^2}}\left[\int_0^{r_a}\frac{\sqrt{{r_a}^2-\rho^2}}{r^2-\rho^2}k(\rho)\rho d\rho + \frac{3\nu}{2-\nu}\frac{\cos 2\theta}{r^2}\int_0^{r_a}\sqrt{{r_a}^2-\rho^2}k(\rho)\rho d\rho\right], \quad r > r_a \tag{S2}$$

where $\nu$ is Poisson's ratio, $\rho$ is the radial coordinate inside the co-seismic zone. Assuming $k(\rho) = -\Delta\tau$ gives the analytical stress distribution for a uniform stress drop:

$$\tau_{zx}(r,\theta) = \frac{2\Delta\tau}{\pi\sqrt{(\frac{r}{r_a})^2-1}}\left\{1-\sqrt{(\frac{r}{r_a})^2-1}\,\tan^{-1}[\frac{1}{\sqrt{(\frac{r}{r_a})^2-1}}] + \frac{\nu\cos 2\theta}{2-\nu}(\frac{r_a}{r})^2\right\}, \quad r > r_a \tag{S3}$$

which decays as $1/\sqrt{r-r_a}$ near the rupture edge and as $1/r^3$ in the far field.

We verify this analytical solution numerically (Figure S2): the numerical and analytical solutions agree well, with a relative error of ~35% only near the rupture edge, where the numerical solution approaches the singularity. This discrepancy arises from the discretization of the circular rupture and the imposed slip.

Integrating Equation S3 over the entire plane outside the rupture yields:

$$\int_{r_a}^{\infty}\int_0^{2\pi}\frac{2\Delta\tau}{\pi\sqrt{(\frac{r}{r_a})^2-1}}\left\{1-\sqrt{(\frac{r}{r_a})^2-1}\,\tan^{-1}[\frac{1}{\sqrt{(\frac{r}{r_a})^2-1}}] + \frac{\nu\cos 2\theta}{2-\nu}(\frac{r_a}{r})^2\right\} r d\theta dr = \pi {r_a}^2\Delta\tau \tag{S4}$$

which confirms the Zheng & Rice (1998) conservation for the shear-stress component resolved along the slip direction.

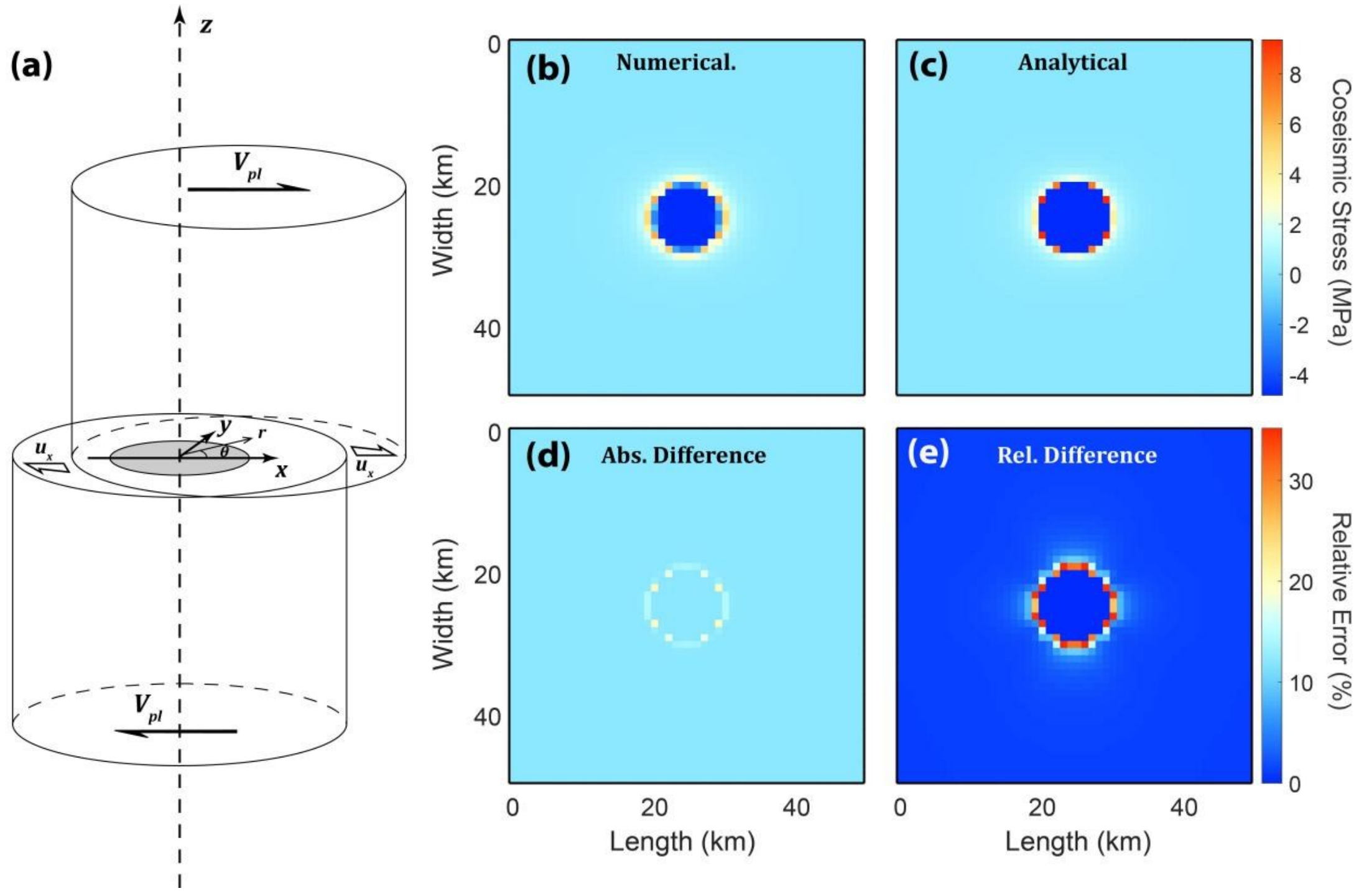

**Figure S2.** Synthetic tests for circular crack model. (a) Pure shear circular crack model. Here, $\boldsymbol{u_x}$ is the shear slip. (b) Spatial distribution of co-seismic stress calculated by numerical simulation. (c) Spatial distribution of co-seismic stress calculated by analytical solution. (d) Stress differences between analytical and numerical results. (b), (c), and (d) share an identical color scale. (e) Relative error between analytical and numerical results for co-seismic stress.

## Text S3

In this section, we demonstrate a theoretical proof to show that the CRF conservation of Section 2.4 extends to a fault in an elastic half-space bounded by a traction-free surface. We consider an arbitrary infinite plane $S$ in a homogeneous, isotropic, elastic half-space (Figure S3). Here $S_f$ denotes the free surface, $S_\infty$ the boundary at infinity, and $S^+$ and $S^-$ the two internal boundaries, which together divide the half-space into two regions $V^+$ and $V^-$.

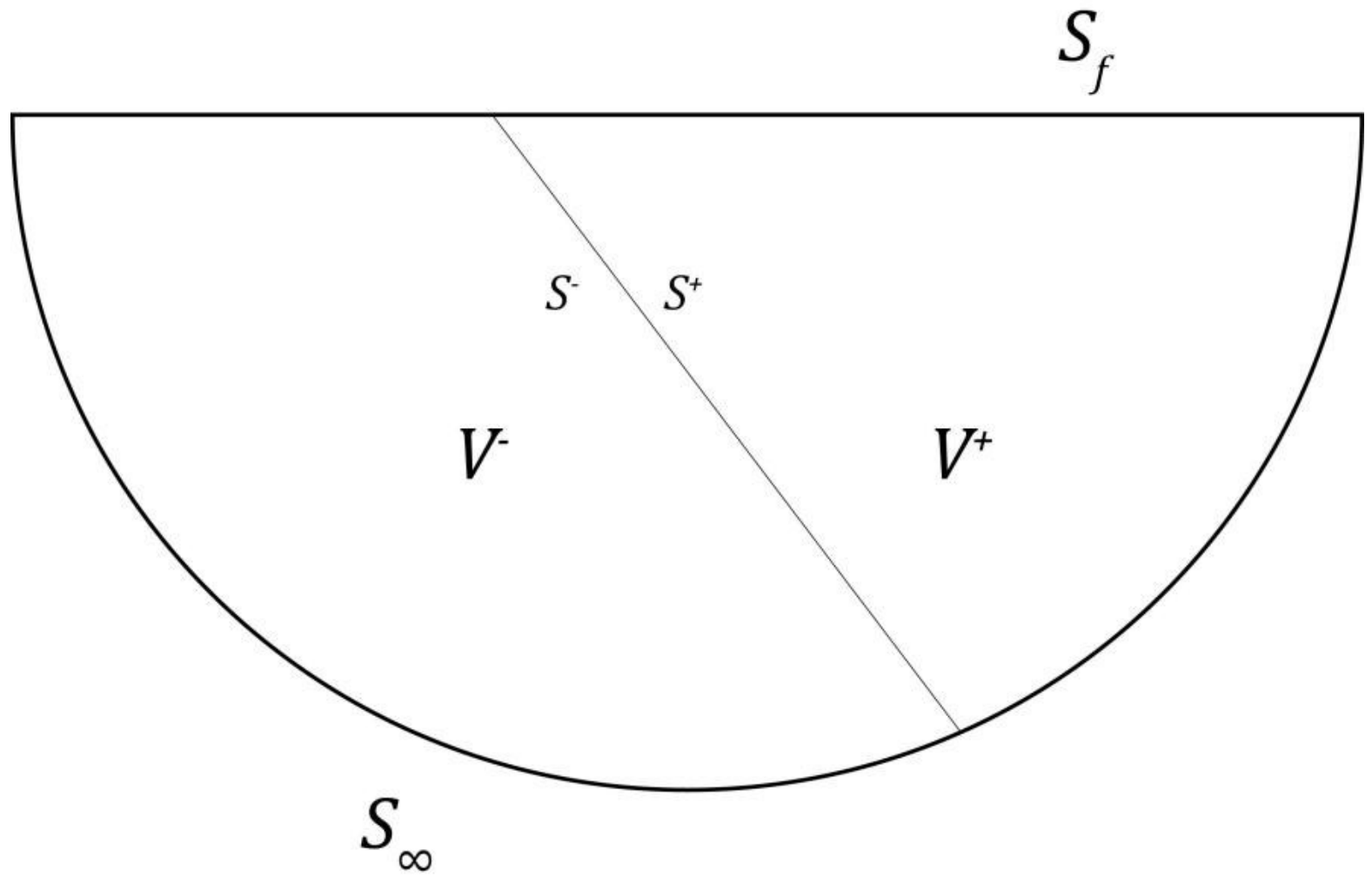

**Figure S3.** Semi-infinite fault model in a half-space.

Consider two purely static force systems: (1) $\boldsymbol{u}(\boldsymbol{x}), \boldsymbol{\sigma}(\boldsymbol{u})$; (2) $\boldsymbol{v}(\boldsymbol{x}), \boldsymbol{\sigma}(\boldsymbol{v})$. In system 1, $\boldsymbol{u}$ and $\boldsymbol{\sigma}$ are the displacement and stress fields produced by a dislocation on an arbitrary finite region of the interface $S$:

$$[\boldsymbol{u}(\boldsymbol{x})] = \delta\boldsymbol{u}(\boldsymbol{x}),\ \boldsymbol{x} \in S \tag{S5}$$

In system 2, let the displacement field be:

$$\boldsymbol{v}(\boldsymbol{x}) = \begin{cases} \boldsymbol{v}^+, \boldsymbol{x} \in V^+ \\ \boldsymbol{v}^-, \boldsymbol{x} \in V^- \end{cases} \tag{S6}$$

where $\boldsymbol{v}^+$ and $\boldsymbol{v}^-$ are constant vectors, so that plane $S$ undergoes a uniform dislocation $[\boldsymbol{v}(\boldsymbol{x})] = \boldsymbol{v}^+ - \boldsymbol{v}^- = \delta v \hat{\boldsymbol{e}}_i$, with $\hat{\boldsymbol{e}}_i$ the unit tangential vector on plane $S$.

At any point within the regions $V^+$ and $V^-$ in system 2,

$$\varepsilon_{ij}(\boldsymbol{v}) = \frac{1}{2}[v_{i,j}(\boldsymbol{x}) + v_{j,i}(\boldsymbol{x})] = 0 \tag{S7}$$

$$\sigma_{ij}(\boldsymbol{v}) = C_{ijkl}\,\varepsilon_{kl}(\boldsymbol{v}) = 0 \tag{S8}$$

where $\varepsilon_{ij}$ is the strain field and $C_{ijkl}$ is the elastic stiffness tensor. This solution simultaneously satisfies the free-surface condition ($\boldsymbol{\sigma} = 0$), the condition at infinity ( $\boldsymbol{\sigma} \to 0$ ), and the internal interface condition ($\boldsymbol{\sigma}(\boldsymbol{v})$continuous); by the uniqueness theorem, it is therefore the solution for system 2.

By Betti's first reciprocity theorem:

$$\iiint_{V^+ + V^-} (\boldsymbol{f}(\boldsymbol{x}) - \rho\ddot{\boldsymbol{u}}(\boldsymbol{x}))\boldsymbol{v}(\boldsymbol{x})dV + \oint\!\!\!\oint_{S_f + S_\infty + S^+ + S^-} \boldsymbol{T}(\boldsymbol{u}, \widehat{\boldsymbol{n}})\boldsymbol{v}(\boldsymbol{x})dS$$

$$= \iiint_{V^+ + V^-} (\boldsymbol{g}(\boldsymbol{x}) - \rho\ddot{\boldsymbol{v}}(\boldsymbol{x}))\boldsymbol{u}(\boldsymbol{x})dV + \oint\!\!\!\oint_{S_f + S_\infty + S^+ + S^-} \boldsymbol{T}(\boldsymbol{v}, \widehat{\boldsymbol{n}})\boldsymbol{u}(\boldsymbol{x})dS \qquad \text{(S9)}$$

its static form (with body force $\boldsymbol{f} = \boldsymbol{g} = 0$ and acceleration $\ddot{\boldsymbol{u}} = \ddot{\boldsymbol{v}} = 0$ ) becomes:

$$\oint\!\!\!\oint_{S_f + S_\infty + S^+ + S^-} \boldsymbol{T}(\boldsymbol{u}, \widehat{\boldsymbol{n}})\boldsymbol{v}(\boldsymbol{x})dS = \oint\!\!\!\oint_{S_f + S_\infty + S^+ + S^-} \boldsymbol{T}(\boldsymbol{v}, \widehat{\boldsymbol{n}})\boldsymbol{u}(\boldsymbol{x})dS \qquad \text{(S10)}$$

where the traction $\boldsymbol{T}$ satisfies:

$$\begin{cases} \boldsymbol{T}(\boldsymbol{u}, \widehat{\boldsymbol{n}}) = \widehat{\boldsymbol{n}} \cdot \boldsymbol{\sigma}(\boldsymbol{u}) \\ \boldsymbol{T}(\boldsymbol{v}, \widehat{\boldsymbol{n}}) = \widehat{\boldsymbol{n}} \cdot \boldsymbol{\sigma}(\boldsymbol{v}) \end{cases} \qquad \text{(S11)}$$

and $\hat{n}$ denotes the unit normal on the interface. Since $\boldsymbol{T}(\boldsymbol{v}, \widehat{\boldsymbol{n}}) = \widehat{\boldsymbol{n}} \cdot \boldsymbol{\sigma}(\boldsymbol{v}) = 0$ for Equation S8, we have:

$$\oint\!\!\!\oint_{S_f + S_\infty + S^+ + S^-} \boldsymbol{T}(\boldsymbol{u}, \widehat{\boldsymbol{n}})\boldsymbol{v}(\boldsymbol{x})dS = 0 \qquad \text{(S12)}$$

On the free surface $S_f$, $\boldsymbol{T}(\boldsymbol{u}, \widehat{\boldsymbol{n}}) = 0$, so

$$\oint\!\!\!\oint_{S_f} \boldsymbol{T}(\boldsymbol{u}, \widehat{\boldsymbol{n}})\boldsymbol{v}(\boldsymbol{x})dS = 0 \qquad \text{(S13)}$$

As for the boundary at infinity $S_\infty$, the static displacement field of a dislocation source decays as $r^{-2}$ in the far field (Aki & Richards, 1980), so its stress field $\boldsymbol{\sigma}(\boldsymbol{u})$ decays as $r^{-3}$; hence,

$$\oint\!\!\!\oint_{S_\infty} \boldsymbol{T}(\boldsymbol{u}, \widehat{\boldsymbol{n}})\boldsymbol{v}(\boldsymbol{x})dS = 0 \qquad \text{(S14)}$$

On the internal boundaries $S^-$ and $S^+$, let the unit normal $\widehat{\boldsymbol{n}}$ be $\widehat{\boldsymbol{e}}_{\boldsymbol{n}}$ and $-\widehat{\boldsymbol{e}}_{\boldsymbol{n}}$, respectively. Since $\boldsymbol{\sigma}(\boldsymbol{u})$ is continuous across them,

$$\oint\!\!\!\oint_{S^+ + S^-} \boldsymbol{T}(\boldsymbol{u}, \widehat{\boldsymbol{n}})\boldsymbol{v}(\boldsymbol{x})dS = \oint\!\!\!\oint_S -\widehat{\boldsymbol{e}}_{\boldsymbol{n}} \cdot \boldsymbol{\sigma}(\boldsymbol{u})\boldsymbol{v}^+ dS + \oint\!\!\!\oint_S \widehat{\boldsymbol{e}}_{\boldsymbol{n}} \cdot \boldsymbol{\sigma}(\boldsymbol{u})\boldsymbol{v}^- dS$$

$$= -\oint\!\!\!\oint_S \widehat{\boldsymbol{e}}_{\boldsymbol{n}} \cdot \boldsymbol{\sigma}(\boldsymbol{u})\delta v \widehat{\boldsymbol{e}}_{\boldsymbol{i}} dS = 0 \qquad (S15)$$

That is,

$$\delta v \oint\!\!\!\oint_S \widehat{\boldsymbol{e}}_{\boldsymbol{n}} \cdot \boldsymbol{\sigma}(\boldsymbol{u})\widehat{\boldsymbol{e}}_{\boldsymbol{i}} dS = \delta v \oint\!\!\!\oint_S \tau_{ni}(\boldsymbol{u})dS = 0 \qquad \text{(S16)}$$

Since $\delta v \neq 0$, $\oiint_S \tau_{ni}(\boldsymbol{u})dS = 0$. Hence the CRF also holds in a half-space.

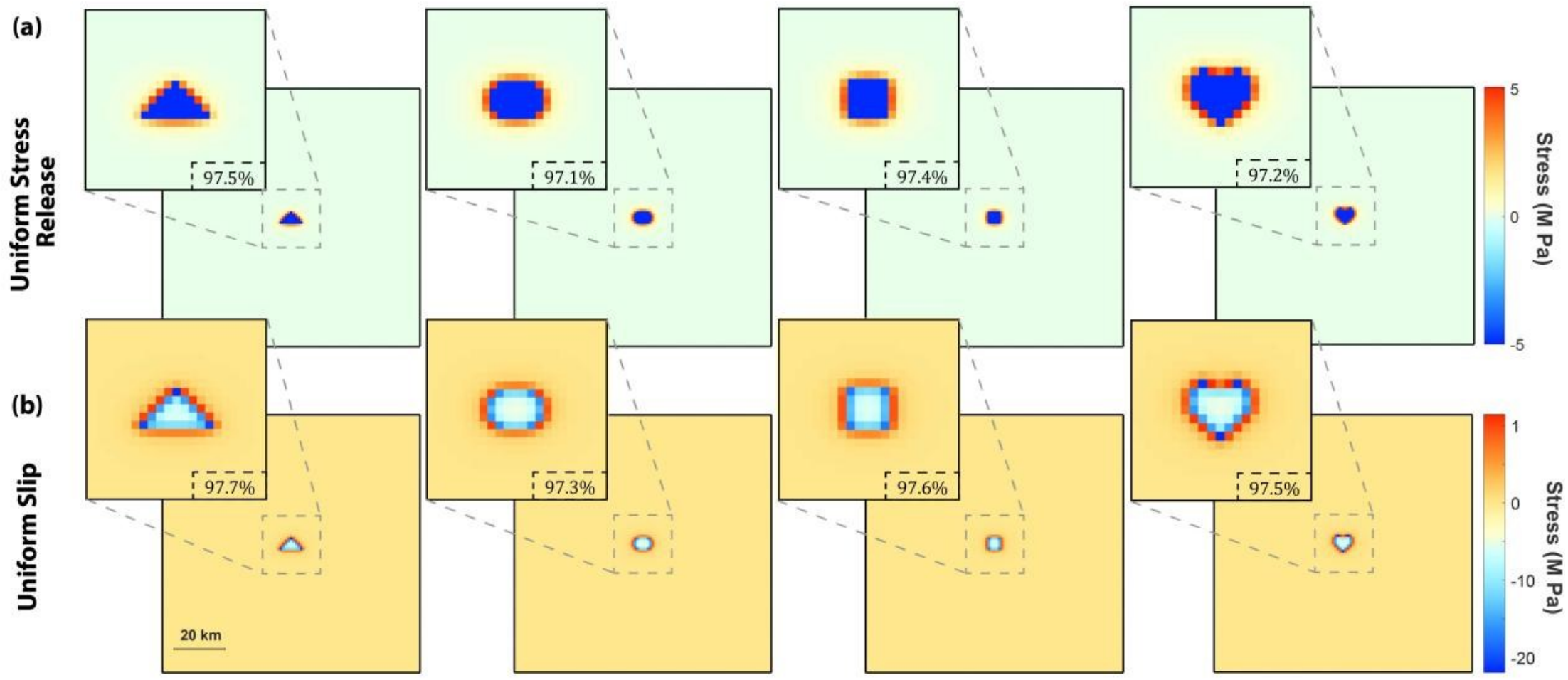

**Figure S4.** Synthetic tests for conservative rupture forces under different rupture patterns. (a) Stress distribution for different rupture patterns under uniform stress drop conditions. (b) Stress distribution for different rupture patterns under uniform slip conditions. The percentage in the inset diagram represents the conservative force ratio $\boldsymbol{R_F}$.